\documentclass{emulateapj}

\usepackage{color}

\newcommand{\simgt}{\lower.5ex\hbox{$\; \buildrel > \over \sim \;$}}
\newcommand{\simlt}{\lower.5ex\hbox{$\; \buildrel < \over \sim \;$}}

\slugcomment{24, February, 2010, ApJ accepted}

\shorttitle{Suzaku Observation of Abell~1689}
\shortauthors{Kawaharada et al.}

\begin{document}

\title{Suzaku Observation of Abell 1689: Anisotropic Temperature 
and Entropy Distributions Associated with the 
Large-Scale Structure}

\author{
Madoka {\sc Kawaharada}\altaffilmark{1}, 
Nobuhiro {\sc Okabe}\altaffilmark{2,3}, 
Keiichi {\sc Umetsu}\altaffilmark{2}, 
Motokazu {\sc Takizawa}\altaffilmark{4}, 
Kyoko {\sc Matsushita}\altaffilmark{5}, 
Yasushi {\sc Fukazawa}\altaffilmark{6}, 
Takashi {\sc Hamana}\altaffilmark{7}, 
Satoshi {\sc Miyazaki}\altaffilmark{7}, 
Kazuhiro {\sc Nakazawa}\altaffilmark{8}, 
{\sc and}
Takaya {\sc Ohashi}\altaffilmark{9} 
}  

\altaffiltext{1}{RIKEN (The Institute of Physical and Chemical Research), 2-1 Hirosawa, Wako, Saitama 351-0198, Japan; kawahard@crab.riken.jp.}
\altaffiltext{2}{Institute of Astronomy \& Astrophysics, Academia Sinica, P. O. Box 23-141 Taipei 10617, Taiwan}
\altaffiltext{3}{Astronomical Institute, Tohoku University, Aramaki, Aoba-ku, Sendai 980-8578, Japan}
\altaffiltext{4}{Department of Physics, Yamagata University, Yamagata 990-8560, Japan}
\altaffiltext{5}{Department of Physics, Tokyo University of Science, 1-3 Kagurazawa, Shinjuku-ku, Tokyo 162-8601, Japan}
\altaffiltext{6}{Deparment of Physical Science, Hiroshima University, 1-3-1 Kagamiyama, Higashi-Hiroshima, Hiroshima 739-8526, Japan}
\altaffiltext{7}{National Astronomical Observatory of Japan, Mitaka, Tokyo 181-8588, Japan}

\altaffiltext{8}{Department of Physics, The University of Tokyo, 7-3-1 Hongo,  Bunkyo-ku,  Tokyo 113-0033, Japan}

\altaffiltext{9}{Department of Physics, Tokyo Metropolitan University, 1-1 Minami-Osawa, Hachioji, Tokyo 192-0397, Japan }

\begin{abstract}

We present results of new, deep \emph{Suzaku} X-ray observations ($160$\,ks)
of the intracluster medium (ICM) in Abell~1689 out to
its virial radius, combined with complementary data sets of the projected
galaxy distribution obtained from the \emph{SDSS} catalog and
the projected mass distribution from our recent comprehensive
weak and strong lensing analysis of \emph{Subaru/Suprime-Cam} and \emph{HST/ACS} observations.
Faint X-ray emission from the ICM around the virial radius ($r_{\rm
 vir}\sim15\farcm6$) is detected at $4.0 \sigma$ significance, thanks
 to low and stable X-ray background of \emph{Suzaku}.
The \emph{Suzaku} observations reveal anisotropic gas temperature and entropy
distributions in cluster outskirts of $r_{500} \simlt r \simlt r_{\rm vir}$
correlated with large-scale structure of galaxies in a photometric
redshift slice around the cluster.
The high temperature ($\sim5.4~$keV) and entropy region in the northeastern (NE) outskirts is apparently
connected to an overdense filamentary structure of galaxies outside the cluster.
The gas temperature and entropy profiles in the NE direction
are in good agreement, out to the virial radius,
with that expected from a recent \emph{XMM-Newton} statistical study
and with an accretion shock heating model of the ICM, respectively.
To the contrary, the other outskirt regions in contact with low density void environments
have low gas temperatures ($\sim 1.7$\,keV) and entropies, deviating from hydrostatic equilibrium.
These anisotropic ICM features associated
 with large-scale structure environments suggest
that the thermalization of the ICM occurs
faster along overdense filamentary structures than along low-density void regions.
We find that the ICM density distribution is fairly isotropic,
with a three-dimentional density slope of  $-2.29\pm0.18$ in the radial range of
$r_{2500} \simlt r \simlt r_{500}$,
and with $-1.24_{-0.56}^{+0.23}$ in  $r_{500} \simlt r \simlt r_{\rm vir}$,
which however is significantly shallower than the Navarro, Frenk, \& White
universal matter density profile in the outskirts, $\rho\propto r^{-3}$.
A joint X-ray and lensing analysis shows that the hydrostatic
mass is lower than spherical lensing one ($\sim60-90\%$) but comparable to a triaxial
 halo mass within errors, at intermediate radii of $0.6r_{2500} \simlt r \simlt 0.8r_{500}$.
On the other hand, the  hydrostatic mass within $0.4r_{2500}$ is
 significantly biased as low as $\simlt60\%$, irrespective of mass models.
The thermal gas pressure within $r_{500}$ is, at most, $\sim50$--$60\%$ of the total pressure
to balance fully the gravity of the spherical lensing mass,
and $\sim30$--$40\%$ around the virial radius.
Although these constitute lower limits when one considers the possible halo triaxiality,
these small relative contributions of thermal pressure would require additional sources of pressure,
such as bulk and/or turbulent motions.

\end{abstract}


\section{Introduction}

Clusters of galaxies are the largest self-gravitating systems in the universe,
and X-ray observations have revealed characteristics of intracluster medium (ICM)
which carries $\sim 70$\% mass of known baryons, such as temperature,
ICM mass, and metals therein. However, the ICM emission has been detectable only to 
$\sim 0.6$ times the virial radius $r_{\rm vir}$ \citep[e.g.][]{pratt-2007} 
due to limited sensitivities of detectors. This means $\sim 80$\% of an entire 
cluster volume has been unexplored in X-rays.

According to
the hierarchical structure formation scenario based on cold dark matter (CDM) paradigm,
mass accretion flows onto clusters, in particular, along filamentary structures, 
are still on-going. 
The ICM in the cluster outskirts, therefore, 
is expected to be sensitive to the structure formation and cluster evolution.
However, since the X-ray data around the virial radius of most clusters
were unavailable,
we have not yet known the ICM characteristics in detail,  
such as density, temperature, pressure and entropy. 
The ICM around the virial radius is a fascinating frontier for X-ray observations.


In the era of {\it Suzaku} \citep{mitsuda-2007}, detection of the ICM beyond 0.6 $r_{\rm vir}$ has become
possible thanks to low and stable particle background of the XIS detectors
\citep{koyama-2007}. In fact, \citet{reiprich-2009} measured temperature profile 
of Abell~2044 out to $r_{\rm 200}$, a radius within which the mean cluster mass density 
is 200 times the cosmic critical density, and found that the profile is in good agreement 
with hydrodynamic simulations. \citet{george-2009} determined radial profiles of
density, temperature, entropy, gas fraction, and mass up to $\sim 2.5\; h_{70}^{-1}$ Mpc
of PKS~0745-191, and found that the temperature profile deviates from that 
expected from the ICM in hydrostatic equilibrium with 
a universal mass density profile as found by Navarro, Frenk \& White
(1996, 1997, hereafter NFW profile).
\citet{bautz-2009} also detected X-ray emission to $r_{\rm 200}$ and found evidence 
for departure from hydrostatic equilibrium 
at radii as small as $r \sim 1.3$ Mpc ($\sim r_{\rm 500}$). 
\citet{fujita-2008} obtained metallicity of the ICM up to the cluster virial radii 
for the first time from a link region between the galaxy clusters Abell~399 and Abell~401.

A gravitational lensing study is complementary to X-ray measurements, 
because lensing observables do not require any assumptions on the cluster dynamical states.
The huge cluster mass distorts shapes of background galaxy images due to
differential deflection of light paths, in a coherent pattern. 
As a result,
multiple images, arcs and Einstein rings caused by the strong gravitational field of a cluster,
appear around the cluster center, which is so-called strong gravitational lensing effect.
It is capable of measuring mass distributions of a cluster central region very well.
Outside the core, weak gravitational lensing analysis, 
which utilizes the coherent distortions in a statistical way,
is a powerful tool to constrain the mass distribution out to the virial radius.
In particular, the {\it Subaru/Suprime-Cam} is the best instrument to conduct the weak lensing analysis
\citep[e.g.][]{miyazaki-2002,hamana-2003,broadhurst-2005b,umetsu-2008,umetsu-2009a,umetsu-2009b, hamana-2009,okabe-2008,okabe-2009,oguri-2009},
 thanks to its high photon-collecting power and high imaging quality, combined with a wide field-of-view.

Since the strong lensing analysis is complementary to weak lensing one,
a joint analysis enables us to determine accurately the cluster mass profile from the center to the virial radius.
Pioneer joint strong and weak lensing studies on Abell~1689 
\citep[][]{broadhurst-2005b,umetsu-2008}
have shown that observed lensing profile is well fitted by an NFW mass profile
with a high concentration parameter 
which is defined as a ratio of the virial radius to the scaled radius.
Therefore, one of the best targets for the study of the ICM out to cluster outskirts is Abell~1689.
Comparisons of X-ray observables with the well-determined lensing mass would allow us to conduct 
a powerful diagnostic of the ICM states, including a stringent test for hydrostatic equilibrium, 
for the first time.
In addition, two mass measurements using X-ray and lensing analysis are of vital importance 
to understand the systematic measurement bias and construct well-calibrated cluster mass,
which will improve the accuracy of determining cosmological parameters using a mass function of galaxy clusters
\cite[e.g. ][]{vikhlinin-2009,okabe-2010,zhang-2010}.

In this paper, we describe the \emph{Suzaku }observations and 
data reduction in \S \ref{section:obs}, and
X-ray background estimation in \S \ref{section:xbkg}.
In \S \ref{section:result}, we show results of spatial and spectral analyses and
obtain radial profiles of temperature, electron density, and entropy. 
In \S \ref{section:dis}, we discuss the hydrostatic equilibrium around the virial radius,
and compare X-ray observables, such as hydrostatic mass, pressure and entropy, 
with lensing mass derived from a joint strong and weak lensing analysis.
We also discuss the ICM characteristics in the outskirts in light of its cosmological environment, 
such as the large-scale structure and low-density voids.
We summarize the conclusions in \S \ref{section:summary}.

In the present work, unless otherwise stated, 
we use $H_0 = 71\; \rm{km\; s^{-1}\; Mpc^{-1}}$ 
($h = H_0 / 100\; \rm{km\; s^{-1}\; Mpc^{-1}} = 0.71$), 
assuming a flat universe with
 $\Omega_{\rm M} = 0.27$. This gives physical scale $1'' = 3.047$ kpc 
at the cluster redshift $z = 0.1832$ \citep{struble-1999}. 
The definition of solar abundance is taken from Anders and Grevesse~(1988). 
Errors are given at the 90~\% confidence level.


\section{Observation and Data Reduction}
\label{section:obs}
\subsection{ \it Observations 
\label{subsection:obs:obs}}

As shown in figure~\ref{fig:mosaic}, we performed four-pointing {\it Suzaku}
observations of Abell~1689, 
named Offset1 (northeast direction), Offset2 (southeast), Offset3
(southwest), and Offset4 (northwest), in the end of July 2008 
with exposure of $\sim 39$ ks for each pointing. 
The pointings were coordinated so that the X-ray emission centroid of Abell~1689 
was located at one corner of each pointing and the entire field-of-view (FOV) of the XIS 
covered the ICM emission up to the virial radius \citep[$15\farcm6 \sim 2.9$ Mpc,][]{umetsu-2008}.
The observation log is summarized in table~\ref{table:observation}. 
All the XIS sensors (XIS0, XIS1 and XIS3) were in normal clocking mode without 
window or burst options throughout the observations.


\subsection{\it XIS data reduction 
\label{subsection:obs:reduction}}
In the present paper, we used the XIS data created through
version 2.2 pipe-line data processing, which is available from the {\it Suzaku} 
ftp site\footnote{ftp://ftp.darts.isas.jaxa.jp/pub/suzaku/ver2.2/}.
We re-processed unscreened XIS event files using {\it Suzaku} software version 11 in HEAsoft 6.6,
together with the calibration database (CALDB) released on 2009 January 9. 
We applied \verb+xiscoord+, \verb+xisputpixelquality+, 
\verb+xispi+, and \verb+xistime+ in this order, and performed the standard
event screenings. Bad pixels were rejected with \verb+cleansis+, 
employing the option of chipcol=SEGMENT.
We selected events with GRADE 0,2,3,4, and 6. Good time intervals (GTI)
were determined by criteria as
 \verb+"SAA_HXD==0+   \verb+&&+ \verb+T_SAA_HXD+\verb+>436 && COR>6 && ELV>5+ \verb+&& DYE_ELV>20 && AOCU_HK_CNT3_NML_P==1 &&+ \verb+Sn_DTRATE<3+ \verb+&& ANG_DIST<1.5"+. 
Detailed procedures of the data processing and screening are the same as those described in {\it The Suzaku Data Reduction Guide}.\footnote{http://ftools.gsfc.nasa.gov/docs/suzaku/analysis/abc/}

Redistribution matrix files (RMFs) of the XIS were produced by 
\verb+xisrmfgen+, and auxillary response files (ARFs) 
by \verb+xissimarfgen+ \citep{ishisaki-2007}. 
As an input to the ARF generator, we prepared an X-ray surface brightness 
profile of Abell~1689 using a double-$\beta$ model 
\citep[sum of two $\beta$ models,][]{king-1962}.
The parameters of the model were determined through a least chi-square fit to
a background-subtracted and vignetting-corrected 
0.5--10 keV surface brightness profile, created from the {\it XMM-Newton} MOS1 image of Abell~1689.
Point sources were detected and removed from the image, 
using  \verb+ewavelet+ task of the SAS software version 8.0.0 
with a detection threshold set at $7\sigma$.
As the background, we used a public MOS1 blank-sky dataset which is available from 
the {\it XMM-Newton} website\footnote{http://xmm2.esac.esa.int/external/xmm\_sw\_cal/background/\\blank\_sky.shtml}. 
Among the blank-sky files, we chose data with hydrogen column densities 
between $1.6 \times 10^{20}$ cm$^{-2}$ and $2.0 \times 10^{20}$ cm$^{-2}$, 
to match the hydrogen column density of Abell~1689, $1.82 \times 10^{20}$ cm$^{-2}$ \citep[][weighted average over $1^{\circ}$ cone radius]{dickey-1990}. 
The total exposure of the selected MOS1 blank-sky data is 125 ks. 
The resulting  parameters of double-$\beta$ model are 
($r_{\rm c}$, $\beta$)$=$($82.2 \pm 11.1$ kpc, $0.97 \pm 0.17$) 
for the narrower component and 
($183.4 \pm 12.3$ kpc, $0.70 \pm 0.01$) 
for the wider component, 
where $r_{\rm c}$ is the core radius. 
Absorption below 2 keV, caused by a carbon-dominated contamination material on the XIS 
optical blocking filters, is included in the ARFs.
In the calculation of contamination by \verb+xissimarfgen+, 
differences of the contaminant thickness among the XIS sensors are taken into
account, along with its radial dependences and secular changes \citep{koyama-2007}.

Non X-ray background (NXB) of each XIS sensor was created 
using \verb+xisnxbgen+, which sorts spectra of night-Earth observations according to
the geomagnetic cut-off-rigidity (COR) and makes an averaged  spectrum
weighted by residence times for which Suzaku stayed in each COR interval
\citep{tawa-2008}. 
Integrated period for the NXB creation is $\pm 150$ days 
of the Abell~1689 observations. 
We assumed systematic errors (90\% confidence limit) of 6.0\% and 12.5\% 
for the NXBs \citep{tawa-2008}
from XIS-FI \citep[front illuminated CCD, see ][XIS0 and XIS3 in this paper]{koyama-2007} and XIS-BI 
\citep[backside illuminated CCD, see][XIS1 in this paper]{koyama-2007}, 
respectively, and added them in quadrature to the
corresponding statistical errors.
In the next section, we estimate the X-ray background (XRB) around Abell~1689 
which is crucial to detect weak signals from the ICM around the virial radius.

\section{Estimation of X-ray Background}
\label{section:xbkg}
\subsection{\it Selection of blank-sky observations
\label{subsection:bgd:sel}}

In order to estimate the X-ray background (XRB), we selected the nearest two blank-sky 
{\it Suzaku} observations to Abell~1689; 
one is Q1334-0033 (Observation ID = 702067010, hereafter Q1334) 
which is $6^{\circ}.33$ offset from Abell~1689,
and the other is NGC4636\_GALACTIC\_1 (Observation ID = 802039010, hereafter N4636\_GAL)
which is $8^{\circ}.68$ offset from Abell~1689.
The exposures are 12 ks and 35 ks for Q1334 and NGC4636\_GAL, respectively,  
after the event processing and screening which are the same as those described in 
\S~\ref{subsection:obs:reduction}.

The hydrogen column densities \citep[][weighted average over $1^{\circ}$ cone radius]{dickey-1990} 
of Abell~1689, Q1334, and N4636\_GAL, are $1.82 \times 10^{20}$ cm$^{-2}$,  
$2.02 \times 10^{20}$ cm$^{-2}$, and $1.96 \times 10^{20}$ cm$^{-2}$, respectively. 
They are in good agreement with one another within 11\%, suggesting the same level of
X-ray background in soft X-ray band. Next we checked the count rates obtained 
from ROSAT All Sky survey (RASS) which are available via the NASA's HEASARC 
website\footnote{http://heasarc.gsfc.nasa.gov/cgi-bin/Tools/xraybg/xraybg.pl}.
Since the RASS count rates in the direction of Abell~1689 is contaminated by its own ICM emission,
we took an annular region around Abell~1689 of which inner and outer radii are $0^{\circ}.5$ 
and $1^{\circ}.0$, respectively. The RASS count rates of the two blank-sky fields 
were defined as circular regions of $1^{\circ}$ radius.
As shown in table~\ref{table:rass-cnt}, the count rates of Q1334 and N4636\_GAL 
match those of the annular Abell~1689 field within errors in all the energy bands,
except for count rates of 1/4 keV and 3/4 keV bands in the N4636\_GAL field,   
which are significantly higher than the corresponding rates in the Abell~1689 field.  
Therefore, the cosmic X-ray background (CXB) of Abell~1689 is considered to be represented 
well by both Q1334 and N4636\_GAL observations, while Galactic foreground emission (GFE) 
should be represented only by the Q1334 observation. Thus, in the following spectral analyses, 
we ignored spectra of N4636\_GAL below $1.5$ keV.

\subsection{\it Spectral analysis of the blank-sky observations
\label{subsection:bgd:ana}}
\begin{figure}
  \begin{center}
   \includegraphics[width=0.28\textwidth,angle=-90,clip]{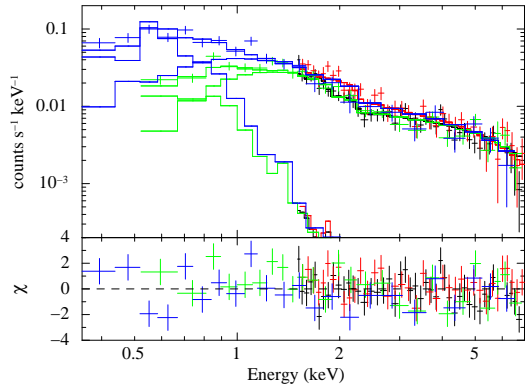} 
  \end{center}
  \caption{NXB-subtracted XIS spectra of N4636\_GAL and Q1334, 
fitted with the X-ray background model described in \S~\ref{subsection:bgd:ana}.   
Black and red are 1.5--7.0 keV XIS-FI (averaged over XIS0 and XIS3) and 1.5--7.0 keV
XIS-BI spectra of N4636\_GAL, respectively.  
Green and blue are 0.5--7.0 keV XIS-FI and 0.3--7.0 keV XIS-BI spectra of Q1334,
respectively. 
}
\label{fig:bgd_fit}
\end{figure}

Using XSPEC12 version 12.5.0ac, we fitted NXB-subtracted XIS-FI (averaged over XIS0 and XIS3) and
XIS-BI spectra of Q1334 and N4636\_GAL simultaneously with a model which represents
the CXB and GFE emissions. RMFs and ARFs were created for each observations by \verb+xisrmfgen+
and \verb+xissimarfgen+, respectively. The input image for the ARFs is uniform emission
over a circular region of $20'$ radius. 
As the CXB model, we used a power law with a fixed photon 
index of 1.41 \citep{kushino-2002} and a free normalization.
To represent the GFE, we employed a model of \citet{henley-2008}, 
which consists of three {\it apec} \citep{smith-2001} components; 
a non-absorbed 0.08 keV component, an absorbed 
0.11 keV one, and an absorbed 0.27 keV one, representing the emission from
the Local Bubble (LB), cool halo, and  hot halo, respectively. 
The three  {\it apec} temperatures were all fixed. 
The hydrogen column density was fixed to the Galatic values of $2.02 \times 10^{20}$~cm$^{-2}$
and $1.96 \times 10^{20}$~cm$^{-2}$ for Q1334 and N4636\_GAL, respectively. 
The metal abundance and redshift were also fixed to one solar and zero, respectively. 
Relative normalizations of the two halo components were tied as 
cool halo : hot halo $= 1 : 0.24$ \citep{henley-2008}.
Thus, the GFE model has only two free parameters; normalizations of the LB and halo. 
As shown in figure~\ref{fig:bgd_fit},  this CXB+GFE model
reproduces the blank-sky spectra successfully. 
The fitting result is summarized in table~\ref{table:fit-bgd}.
The 2.0--10.0 keV surface brightness of the CXB component is
$(6.3 \pm 0.2 \pm 0.6) \times 10^{-8}$ erg s$^{-1}$ cm$^{-2}$ sr$^{-1}$
(90\% statistical and systematic errors), 
where the systematic error, 10\%, refers to Appendix 1 of \citet{nakazawa-2009}. 
This is consistent with the average ASCA result of absolute CXB intensity
\citep{kushino-2002}, 
$(6.38 \pm 0.07 \pm 1.05) \times 10^{-8}$ erg s$^{-1}$ cm$^{-2}$ sr$^{-1}$ 
(90\% statistical and systematic errors).


\section{Analysis and Results}
\label{section:result}
\subsection{\it X-ray surface brightness profile
\label{subsection:ana:sb}}

\begin{figure}
  \begin{center}
    \includegraphics[width=0.3\textwidth,angle=0,clip]{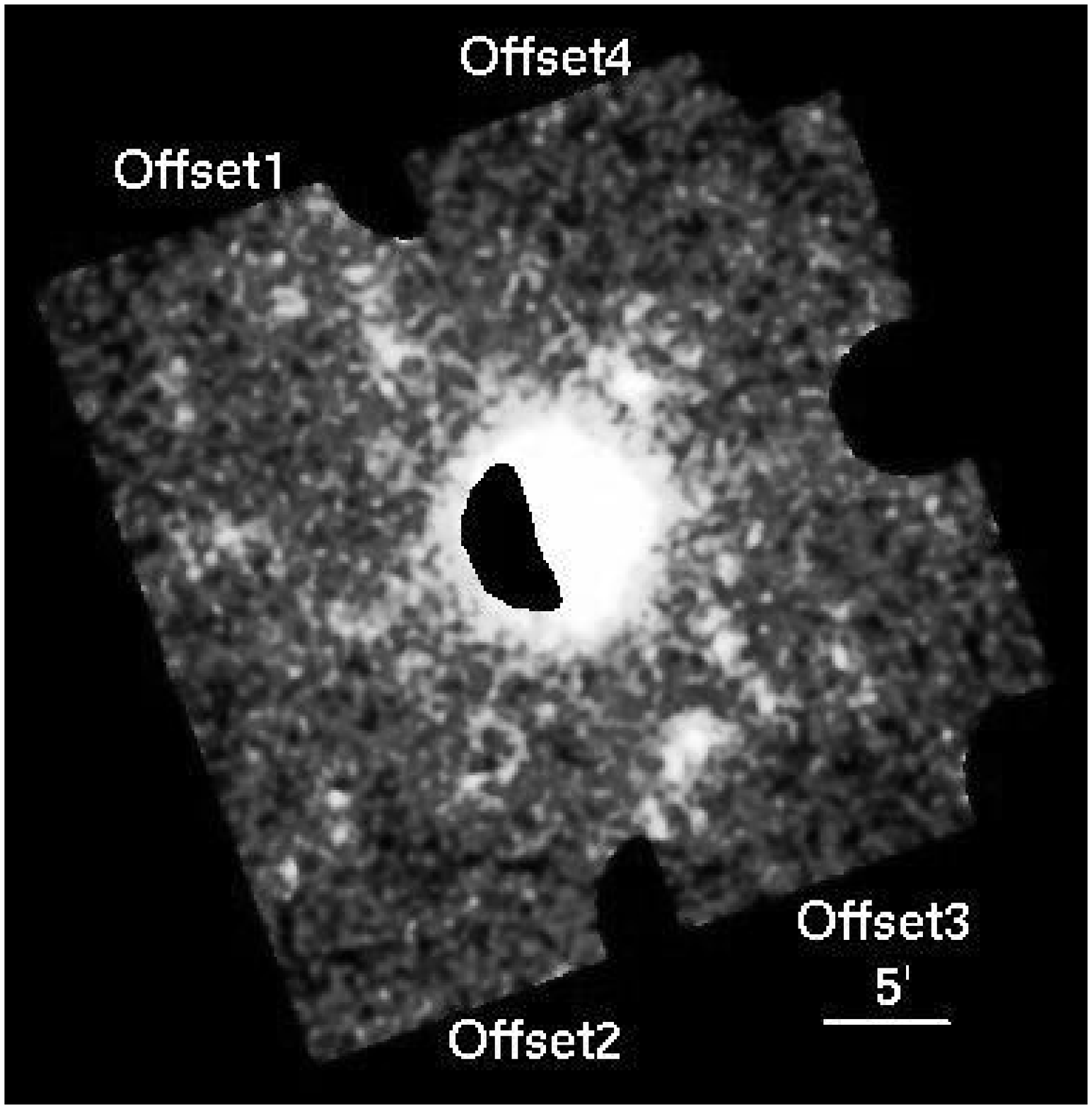} 
   \includegraphics[width=0.315\textwidth,angle=0,clip]{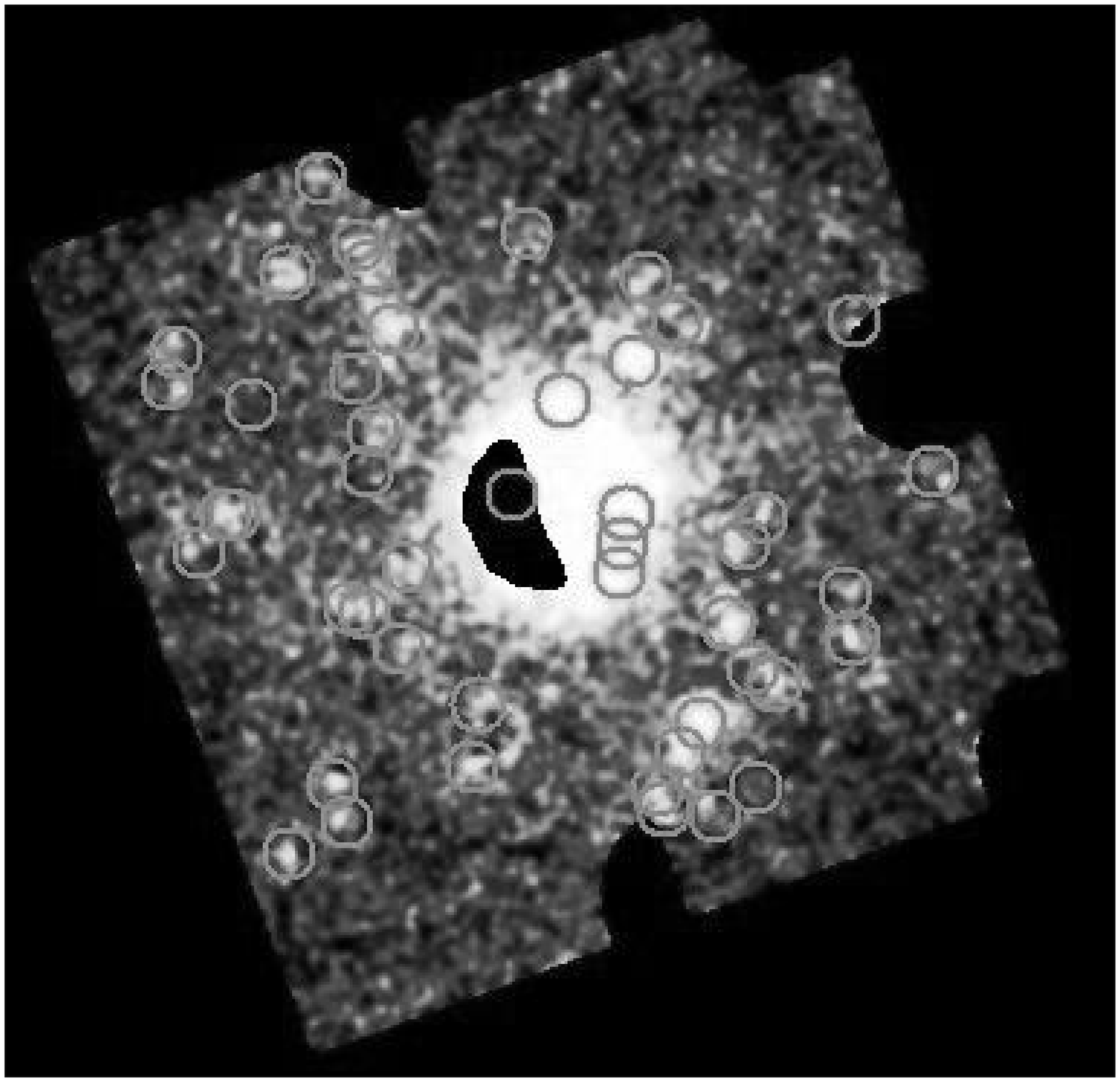} 
  \end{center}
  \caption{(left) A Mosaic image of Abell~1689, created with $0.5-10.0$ keV XIS-FI (XIS0 $+$ XIS3) 
images smoothed with a Gaussian kernel of $\sigma = 10''$. 
Background is not subtracted and vignetting is not corrected. 
CCD regions where $^{55}$Fe calibration sources are irradiated  (\cite{koyama-2007}) are excluded. 
(right) the same as the left panel, but point-source regions are specified as circles of 
$1'$ radius. See text for details of the point-source detection.
}
\label{fig:mosaic}
\end{figure}

In order to see how far the ICM emission of Abell~1689 is detected, 
we made a surface brightness profile from the 0.5--10 keV 
XIS-FI (XIS0 $+$ XIS3) mosaic image (figure~\ref{fig:mosaic}). The center was
chosen to be ($\alpha$, $\delta$)=($13^{\rm h}11^{\rm m}29.495^{\rm s}, -01^{\circ}20'29.92''$)
in J2000 coordinates as a position of X-ray emission centroid in a 
0.5--10 keV {\it XMM-Newton} MOS1 image.
This position is close to the brightest-cluster galaxy, 
MaxBCG~J197.87292-01.34110~BCG, $2''.4$ ($\sim 7.3$ kpc) offset from the center.
We identified point sources from the MOS1 image using \verb+ewavelet+ task of the SAS 
software version 8.0.0 with a detection threshold set at 7$\sigma$. As to point 
sources outside the MOS1 image, we selected visible sources in a full band image 
of ROSAT PSPC of which flux limit corresponds to 
$\sim 2 \times 10^{-14}$ erg s$^{-1}$ in 2.0--10.0 keV. 
As shown in figure~\ref{fig:mosaic} (right), circular regions ($1'$ in radius) 
around 49 point sources were masked in the XIS-FI mosaic image 
(their positions and flux are summarized in table~\ref{table:pntsrc} 
in Appendix~B).

In figure~\ref{fig:prof_all}, we show 0.5--2 keV and 2--10 keV raw 
(background inclusive, but point-source excluded) 
surface brightness profiles as black crosses. 
We created corresponding 0.5--2 keV and 2--10 keV NXB profiles from 
an NXB mosaic image with \verb+xisnxbgen+. 
The 0.5--2 keV and 2--10 keV NXB profiles are shown in magenta in figure~\ref{fig:prof_all}.
Then, we obtained an XRB mosaic image. 
Based on the CXB and GFE models of Abell~1689 described in \S~\ref{subsection:bgd:ana}, 
we simulated XRB (CXB+GFE) images of the four offset observations 
using \verb+xissim+ with exposures 10 times longer than those of actual observations. 
The 0.5--2 keV and 2--10 keV XRB profiles are shown in green in figure~\ref{fig:prof_all}. 
The CXB level of Abell~1689 is not the same as that of the blank-sky fields because
point sources are excluded in Abell~1689. 
Therefore, intensity of the CXB model when the flux is integrated to our
flux limit, 
$2 \times 10^{-14}$ erg s$^{-1}$ in 2.0--10.0 keV, 
was calculated from equation 6 of \cite{kushino-2002}, 
and used in the simulation. 
The resulting CXB intensity of Abell~1689 is 65\% of the absolute CXB intensity. 

%
%


Finally, we obtained a background-subtracted profile 
(black $-$ magenta $-$ green in figure~\ref{fig:prof_all}), 
as shown in red in figure~\ref{fig:prof_all}. 
In the error bars in this background-subtracted profile (red crosses), 
1$\sigma$ uncertainties of the XRB and NXB are added in 
quadrature to the corresponding statistical 1$\sigma$  errors. 
We adopted 3.6\% as the $1 \sigma$ uncertainty of NXB \citep{tawa-2008}, 
while that of XRB was calculated from $1 \sigma$ statistical errors of 
normalizations in the CXB and GFE models, determined from the spectral fitting
of blank sky observations (\S~\ref{subsection:bgd:ana}). 
These are 2.1\% and 15.8\% for the 
CXB and the GFE, respectively, corresponding to 8.5\% of 0.5--10 keV XRB flux. 
As a result, the background-subtracted profiles both in 0.5--2 keV and 2--10 keV bands
extend up to the virial radius ($15\farcm6 \sim 2.9$ Mpc). 
However, as we see in \S~\ref{subsection:ana:syserr}, 
more careful treatment of uncertainties related to a rather broad 
{\it Suzaku} point spread function (PSF), $\sim 2'$ in half power diameter (hereafter HPD),
is needed to evaluate the significance of the signal around the virial radius.

\subsection{\it Significance of the signal
\label{subsection:ana:syserr}}

The signals around the virial radius found in 
the X-ray surface brightness profile (red crosses in figure~\ref{fig:prof_all}) 
have to be tested for systematic uncertainties caused by 
(1) contaminated signals from inner ICM emission, and 
(2) residual signals from the excluded point sources.

As for the item (1), significant amount of signals from inner regions of the ICM is 
expected to contaminate the signal around the virial radius, 
since the PSF of {\it Suzaku} is $\sim 2'$ in HPD
\citep[][]{serlemitsos-2007}.
However, as described in detail in Appendix~A, the simulated 
double-beta profile determined from the 0.5--10 keV {\it XMM-Newton} MOS1 image 
(\S~\ref{subsection:obs:reduction}) is consistent with the observed
profile, and the contaminated signals which originate inside
$10'$ contributes only 22\% of signals outside $10'$.
Therefore, the signal observed around the virial radius cannot be explained by 
the contaminated signals from the inner ICM emission.

\begin{figure}
  \begin{center}
   \includegraphics[width=0.4\textwidth,angle=0,clip]{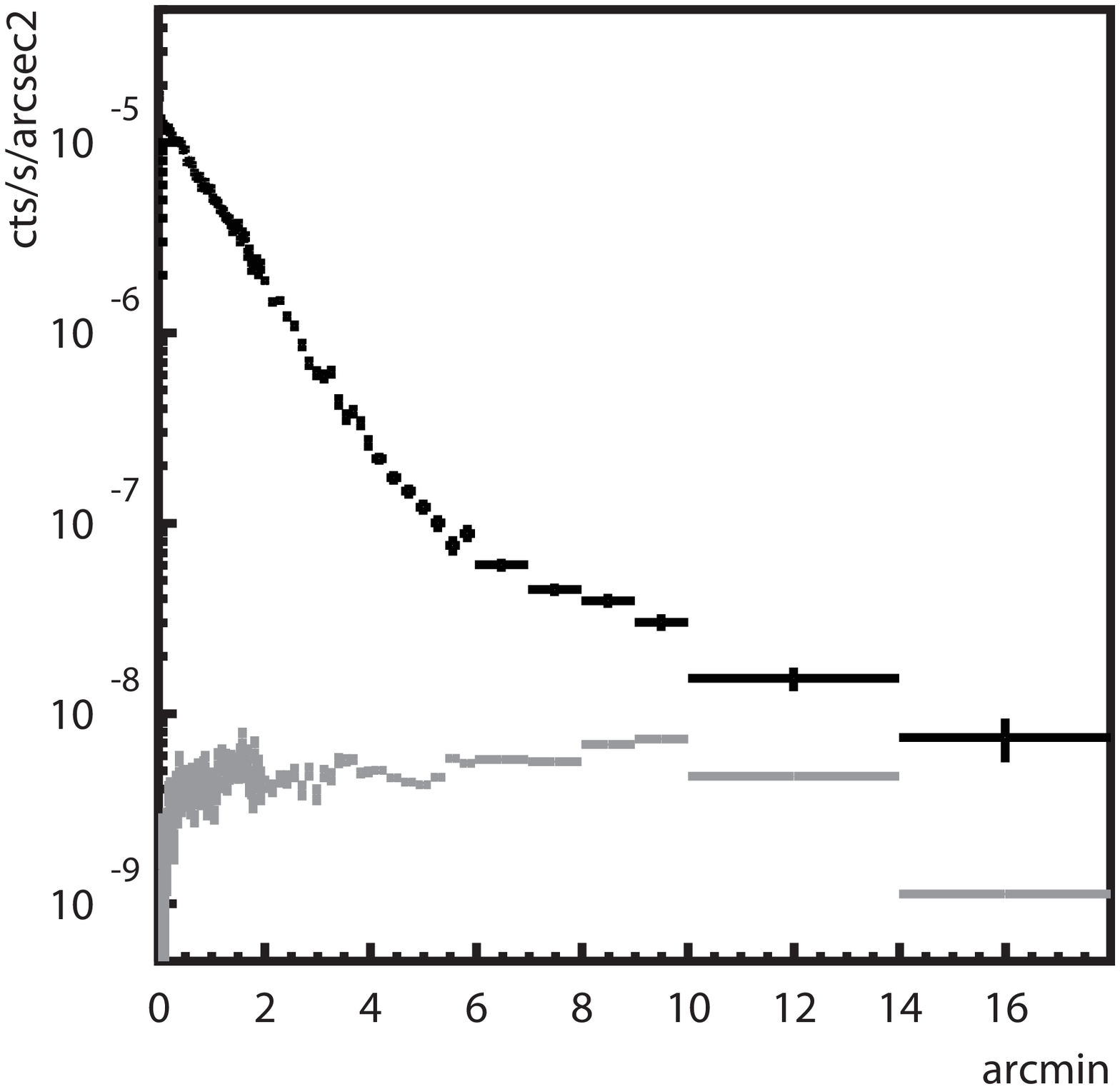} 
   \includegraphics[width=0.4\textwidth,angle=0,clip]{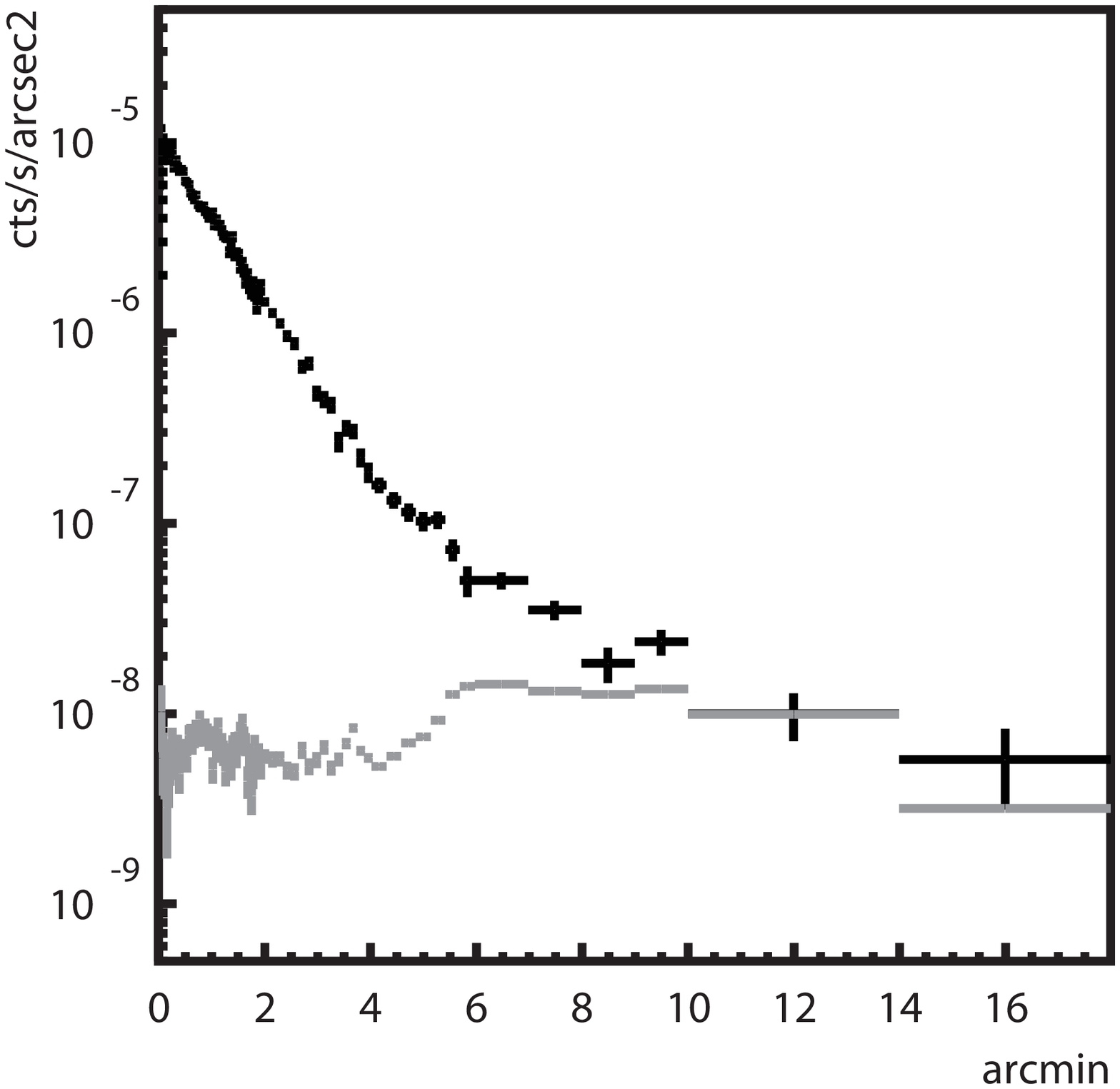} 
  \end{center}
  \caption{
(left) A simulated 0.5--2 keV profile of residual point source signals (gray),
shown with background-subtracted 0.5--2 keV  surface brightness profile of 
XIS-FI averaged over XIS0 and XIS3 (black).
In the error bars in this black profile, 1$\sigma$ uncertainties of the XRB and NXB
are added in quadrature to the corresponding statistical 1$\sigma$  errors. 
(right) the same as the left panel, but for 2--10 keV profiles. 
}
\label{fig:pntsrc_prof}
\end{figure}

The item (2) is also due to the $2'$ HPD of {\it Suzaku}. Although we excluded
circular regions around the point sources, half of signals from point sources is 
expected to escape from the excluded circular regions of $1'$ radius. 
As shown in figure~\ref{fig:pntsrc_prof}, we simulated this residual signals
from point sources. 
Details of the simulation are described in Appendix~B.  
In 0.5--2 keV band, the contribution of the point sources in the
$10\farcm0$--$18\farcm0$ region is only 26\% of the observed profile. Meanwhile, 
the observed 2--10 keV profile can be explained well by the point sources.

By combining the contributions of (1) and (2), we conclude that in 0.5--2 keV band,
these effects explain at most 48\% of the signal around the virial radius ($10\farcm0$--$18\farcm0$),
and the detection is significant at 4.0$\sigma$ level. 
On the other hand, in 2--10 keV range, the signal is not significant. 
This indicates that the detected ICM temperature around the virial radius might be low, 
compared with the central temperature of $\sim 9$ keV \citep{snowden-2008}.
In table~\ref{table:significance}, we show signal significances when X-ray 
surface brightness profiles are evaluated separately for the four pointing directions. 
In the Offset1 direction (northeast), the signal in 2--10 keV band is also detected
with $2.1\sigma$ level, suggesting harder spectrum in this direction.

\subsection{\it Spectral analysis 
\label{subsection:ana:spectral}} 

Since we surely detected the ICM emission out to the virial radius 
from the analysis of X-ray surface brightness, we proceeded to spectral analysis. 
For each pointing (Offset1 through Offset4), we divided the XIS image (figure~\ref{fig:mosaic}) 
to five concentric annular regions centered on the X-ray emission 
centroid (\S~\ref{subsection:ana:sb}); their inner and outer radii are 
$0\farcm0$--$2\farcm0$, $2\farcm0$--$4\farcm0$, $4\farcm0$--$6\farcm0$, 
$6\farcm0$--$10\farcm0$, and $10\farcm0$--$18\farcm0$. 
We have four azimuthal regions at the same distance from the center. 
The 49 point sources were masked by circles of $1'$ radius as we did in obtaining
the X-ray surface brightness profile (\S~\ref{subsection:ana:sb}).
After extracting spectra of the XIS0, XIS1, and XIS3 from each region, 
we averaged XIS-FI (XIS0 and XIS3) spectra. In the $0\farcm0$--$2\farcm0$ and $2\farcm0$--$4\farcm0$
regions, all spectra of Offset1, an XIS-FI spectrum of Offset2, 
and an XIS-BI spectrum of Offset4 are unavailable because regions 
irradiated by the calibration source are removed from the analysis. 

\begin{figure}
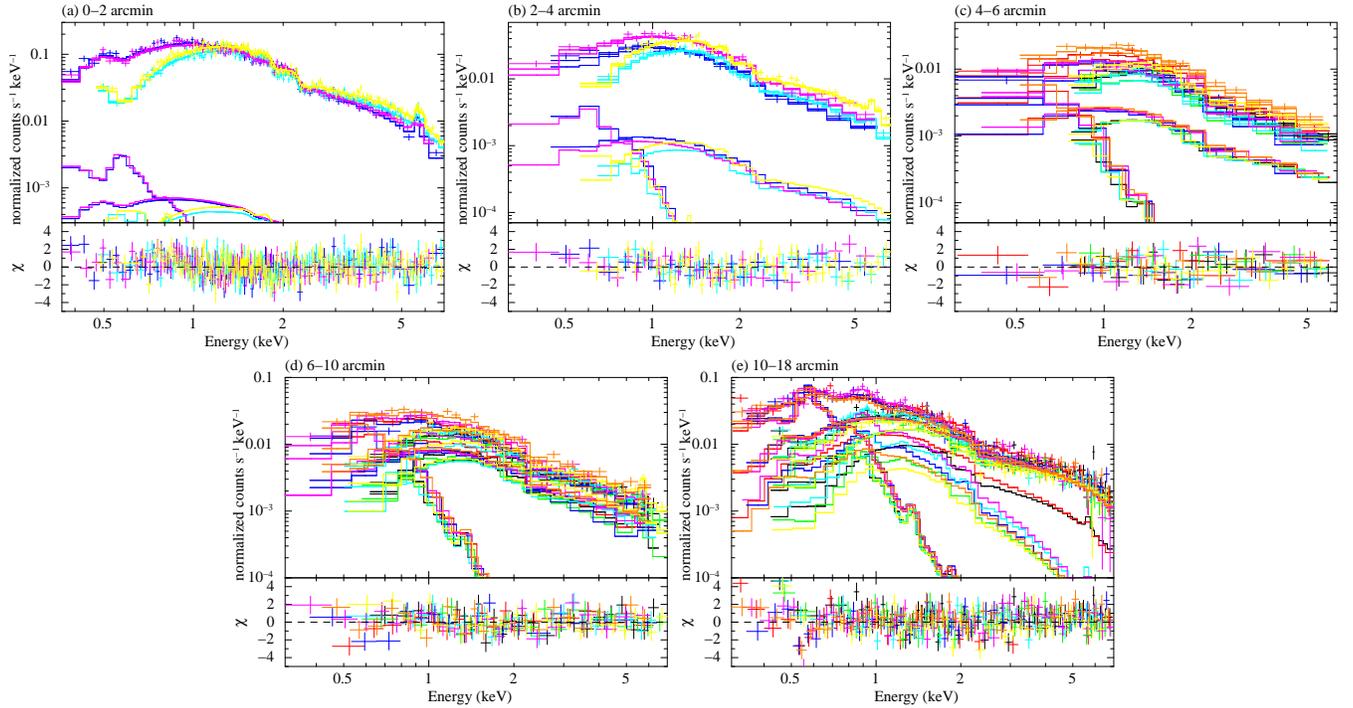

  \begin{center}
   \includegraphics[width=0.26 \textwidth,angle=-90,clip]{./f5a.eps} 
   \includegraphics[width=0.26 \textwidth,angle=-90,clip]{./f5b.eps} 
   \includegraphics[width=0.26 \textwidth,angle=-90,clip]{./f5c.eps} 
   \includegraphics[width=0.26 \textwidth,angle=-90,clip]{./f5d.eps} 
   \includegraphics[width=0.26 \textwidth,angle=-90,clip]{./f5e.eps} 
  \end{center}
  \caption{
NXB-subtracted spectra of (a) $0'$--$2'$, (b) $2'$--$4'$, (c) $4'$--$6'$,
(d) $6'$--$10'$, and (e) $10'$--$18'$ regions, which are extracted from 
Offset1 XIS-FI (black), Offset1 XIS-BI (red), Offset2 XIS-FI (green),
Offset2 XIS-BI (blue), Offset3 XIS-FI (cyan), Offset3 XIS-BI (magenta), 
Offset4 XIS-FI (yellow), and Offset4 XIS-BI (orange). 
These spectra are fitted simultaneously with {\it phabs} $\times$ {\it apec} model, added
to the XRB (GFE and CXB) model (\S~{\ref{subsection:bgd:ana}}). 
Although blank-sky spectra are also fitted simultaneously with the XRB model, 
they are not plotted here for clarity. See text for details of the spectral
analysis (\S~\ref{subsection:ana:spectral}). 
}
\label{fig:spec}
\end{figure}

As shown in figure~\ref{fig:spec}, we used XSPEC12 version 12.5.0ac and included 
the XIS-FI and XIS-BI spectra from the regions
at the same distance, and corresponding NXB and response (RMF and ARF) files created through the procedure
described in \S~\ref{subsection:obs:reduction}. 
In $10\farcm0$--$18\farcm0$ analysis, for example, we 
included four XIS-FI source spectra and four XIS-BI source spectra.
In order to determine the XRB, 
we also included the spectra of blank-sky fields (figure~\ref{fig:bgd_fit}) 
with their NXB and response files (\S~\ref{subsection:bgd:ana}). 
Then, the NXB-subtracted source spectra were fitted with an absorbed thin-thermal emission model 
represented by {\it phabs} $\times$ {\it apec}, 
added to the XRB model (\S~\ref{subsection:bgd:ana}), while the blank-sky spectra were fitted 
with the XRB model simultaneously with the source spectra. 
Relative CXB normalization of Abell~1689 to the blank-fields was 
fixed to 65\% (\S~\ref{subsection:ana:sb}).
Free parameters in the {\it apec} model are
temperature, metal abundance, and normalization. 
The hydrogen column density was fixed to the Galactic value of $1.82 \times 10^{20}$ cm$^{-2}$ 
and redshift was also fixed to 0.1832. 
These free parameters for different pointings (Offset1 through Offset4) were determined separately 
when we saw azimuthal differences, or tied when we obtained azimuthal averages. 
Relative normalization between the XIS-FI and XIS-BI was left free and its deviations from
unity were within $\pm 9$\% in all the regions. 
The fitting result is summarized in table~\ref{table:fit-result}. 
Although the X-ray background model is determined separately for the individual
annular regions, the normalizations of CXB, LB, and Halo are consistent among the regions
within errors.
In this spectral analysis, we did not explicitly include the two uncertainties described 
in \S~\ref{subsection:ana:syserr}, namely, 
contaminated signals from inner ICM emission and residual signals from the excluded point sources.
We estimate their effect on the temperature and electron number density of Abell~1689
in the following section, obtaining radial profiles of these ICM parameters.

\subsection{\it Radial profiles
\label{subsection:ana:prof}}
\subsubsection*{Temperature Profile}

The top panel of figure~\ref{fig:prof} shows radial profiles of gas temperature, 
obtained from the spectral analysis (\S~\ref{subsection:ana:spectral}). 
In figure~\ref{fig:prof} (and following figures of any radial profiles), the center of each
radial bin is defined as emission-weighted center derived from the X-ray surface brightness
obtained in \S~\ref{subsection:ana:sb}; 
$1\farcm 1$, $2\farcm 7$, $4\farcm 8$, $7\farcm 8$, and  $13\farcm4$ for the five
annular regions.  
Although the temperature profile is a projected one obtained by the two-dimensional 
spectral analysis, it coincides well with a three-dimensional temperature profile 
obtained with {\it Chandra} data \citep{peng-2009} in the overlapping region.
The temperature is $\sim 10$ keV within $4\farcm0$ ($\sim 730$ kpc), and
it gradually decreases toward the outskirts down to $\sim 2$ keV around the virial
radius ($r_{\rm vir}\sim15\farcm6 \sim 2.9$ Mpc). 
The decrease of temperature is independent of the azimuthal directions up to $\sim r_{500}\simeq8\farcm7$, 
at which the mean interior density is $500$ times the critical mass density of the universe.
Note that the overdensity radius $r_{500}$ is determined by lensing analysis, as is the case
for $r_{\rm vir}$ (see \S \ref{sec:joint} and eq. (\ref{eq:viral})).
The logarithmic slopes of temperature between the third and forth bins,
 corresponding to $r_{2500} \simlt r \simlt r_{500}$ are 
$\gamma=-d \log T/ d \log r=0.87\pm0.25, 0.71\pm0.30,~\&~ 0.87\pm0.15$ for all, 
Offset1 and Offset234 directions, respectively,
which are consistent within errors with each other.

\begin{figure}
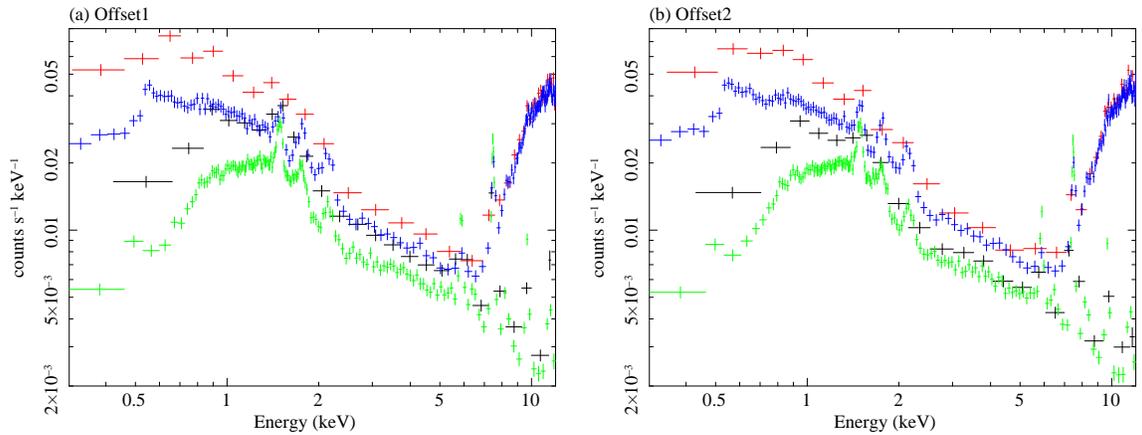

  \begin{center}
   \includegraphics[width=0.32 \textwidth,angle=-90,clip]{./f7a.eps} 
\hspace{0.2cm}
   \includegraphics[width=0.32 \textwidth,angle=-90,clip]{./f7b.eps} 
  \end{center}
  \caption{
(a) Raw XIS-FI (black; averaged over XIS0 and XIS3) and XIS-BI (red) spectra 
    of the Offset1 $10\farcm0$--$18\farcm0$ region, 
    and 
    corresponding  background spectra of XIS-FI (green) and XIS-BI (blue). 
    Each background spectrum was created by adding NXB and XRB spectra.
    The XRB spectrum was simulated using {\it fakeit} command in XSPEC12, 
    incorporating the GFE and CXB models detemined
    by the spectral fitting described in \S~\ref{subsection:ana:spectral}. 
(b) The same as panel (a), but for the Offset2 $10\farcm0$--$18\farcm0$ region.
}
\label{fig:spec-compbgd}
\end{figure}

The temperature variations in direction becomes significant in the outskirts of
 $r_{500} \simlt r \simlt r_{\rm vir}$ : the temperatures for Offset1 ($\sim5.4~$keV) 
and Offset234 ($\sim1.7~$keV) are about half and one-sixth of a mean temperature $10.4$ keV, 
which is an emission-weighted temperature 
of $2\farcm0$--$4\farcm0$ and $4\farcm0$--$6\farcm0$ regions, respectively.
This temperature variations can be seen directly in spectra. 
In figure~\ref{fig:spec-compbgd}, we show the spectra of Offset1 and
Offset2 in $10\farcm0$--$18\farcm0$ region. The signal of Offset1 is 
clearly detected from both XIS-FI and XIS-BI in $2$--$5$ keV range, whereas the signal
of Offset2 is comparable to the background in the same energy range. This means
that the XIS spectra of Offset1 is harder than those of Offset2, resulting in the 
higher temperature in the Offset1 direction.  
The logarithmic slopes of temperature in the outskirts
 corresponding to $r_{500} \simlt r \simlt r_{\rm vir}$ are $\gamma=-d \log T/ d \log r=1.45\pm0.46, 0.18\pm0.41~\&~1.92\pm0.24$ for all, Offset1 and Offset234 directions, respectively.
The slope in Offset234 and all directions are surprisingly steep.

Contaminated signals from the inner ICM emission due to the {\it Suzaku} PSF
may affect the temperature determination. However, in PKS~0745-191 case \citep{george-2009}, the effect is 
at most $\Delta T = -0.8$ keV even in a central region where $>30$\% of photons 
come from the other regions. In the other regions where photon mixing is $<20$\%, 
the temperature uncertainties are $< 0.1$ keV. 
Therefore, in our case of Abell~1689, we expect that the PSF effect is at most $\sim 0.1$ keV 
in the cluster outskirts outside $6'$, which is much smaller than the statistical errors. 
The effect inside $6'$ would be also negligible because our temperature profile
is consistent with the {\it Chandra} result \citep{peng-2009} as described above. 
In order to see the other effect, the residual signals from the excluded point sources
in $10\farcm0$--$18\farcm0$,
we added a power-law of photon index 1.4 into the fit with its normalization 
fixed to the summed escaped flux of the 49 excluded point sources. 
This corresponds to increasing the CXB level by $\sim 15\%$. 
As a result, the effect was as small as $\Delta T \sim -0.3$ keV 
in all the azimuthal directions, which
is also smaller than the statistical errors.

We compared our temperature profiles 
with that based on the scaled temperatures
derived from a sample of 15 clusters by \cite{pratt-2007}. 
We used the mean temperature $10.4$ keV to calculate the scaled profile.
The temperature profile in the Offset1 direction is in good agreement with the scaled temperature one
(dashed line in figure~\ref{fig:prof}(a)), while the temperatures of the other 
regions (Offset2 - Offset4) around the virial radius are lower than the scaled value by a factor of $\sim 2$.

\subsubsection*{Density Profile}

The electron number density profile was calculated from the normalization
parameter of {\it apec} model, defined as
\begin{equation}
Norm = \frac{10^{-14}}{4 \pi (D_{\rm A}(1+z))^2} \int n_{\rm e}n_{\rm H} dV, 
\end{equation}
where $D_{\rm A}$ is the angular size distance to the source in units of cm, 
$n_{\rm e}$ is the electron density in units of cm$^{-3}$, and $n_{\rm H}$ is the 
hydrogen density in units of cm$^{-3}$. Since metal abundance was poorly constrained
from our fitting result (table~\ref{table:fit-result}), we used $n_{\rm H}/n_{\rm e} = 0.852$ 
assuming metal abundance to be 0.3 solar. We calculated a geometrical volume which 
contributes to each two-dimensional region assuming that Abell~1689 is spherically symmetric, 
and obtained projected volume-averaged density for the region. Then, the projected
density profile was deprojected by the following equation
\begin{equation}
\int n_{\rm e,depro}^2 dV_{i} = n_{\rm e, proj}^2 V_{i},
\end{equation}
where $n_{\rm e,depro}$ is the deprojected electron density, 
$n_{\rm e, proj}$ is the projected electron density, and
$V_{i}$ is the volume contributing to $i$-th annular region. We assumed that no emission exists 
outside the $10\farcm0$--$18\farcm0$ region and hence $n_{\rm e,depro} = n_{\rm e, proj}$ 
in the $10\farcm0$--$18\farcm0$ region. 
The result is summarized in table~\ref{table:density}.

The electron number density profile (middle panel of figure \ref{fig:prof})
decreases down to ${\mathcal O}(10^{-5})~{\rm cm^{-3}}$ around the virial radius.
The density profiles are in good agreement with 
a generalized $\beta$ model \citep{vikhlinin-1999} obtained by 
fitting {\it Chandra} data within $9\farcm0$ \citep[][Model 3 in table~9]{peng-2009}. 
In contrast to the temperature profile, 
the electron densities of the four offset directions are consistent 
with one another within errors in any circular annulus.
The slope of the number density is $-2.29\pm0.18$ in the range of $r_{2500} \simlt r \simlt r_{500}$. 
Our result coincides with previous studies around $r_{500}$ : 
$n_e\propto r^{-2.4}$ for the REFLEX-DXL clusters \citep{zhang-2006,zhang-2007}, 
$n_e\propto r^{-2.2\pm0.1}$ for the LoCuSS sample \citep{zhang-2008}, 
$n_e\propto r^{-1.80\pm0.46}$ for the REXCESS sample \citep{croston-2008},
and
$n_e\propto r^{-2.21}$ for 13 clusters observed with \emph{Chandra} \citep{vikhlinin-2006}.
A consistent result, $n_e\propto r^{-2.27\pm0.12}$, has been also obtained with
\emph{Suzaku} in Abell~1795 around $r_{500}$ \citep{bautz-2009}.
In the outer annulus of $r_{500} \simlt r \simlt r_{\rm vir}$, the density slope
becomes flatter to $n_e \propto r^{-1.24\pm0.23}$.   
The same tendency of flattening is seen in a electron density profile of PKS~0745-191 
around $r\simgt2~{\rm Mpc}$ \citep{george-2009}. 
Excess signals in a X-ray surface brightness profile above a $\beta$ model
outside $2~{\rm Mpc}$ seen in 
Abell~1795 \citep{bautz-2009} also indicates the flattening of slope in the
cluster outskirts.
The electron density profile outside $r\sim r_{500}$ is significantly shallower 
than asymptotic matter density slope $\rho\propto r^{-3}$ of 
an NFW profile \citep[][1997]{navrro-1996}, defined by
\begin{eqnarray}
\rho =  \frac{\rho_s}{(r/r_s)(1+r/r_s)^2},
\end{eqnarray}
where $\rho_s$ is the central density parameter and $r_s$ is the scale radius.

Since the electron density profile is consistent with the {\it Chandra} result
within $9\farcm0$ \citep{peng-2009}, we can say that the systematic
effects of the {\it Suzaku} PSF are safely within the statistical errors in this region.
In our $10\farcm0$--$18\farcm0$, however,
48\% of the signal can leak from
the other regions or point sources
as discussed in \S~\ref{subsection:ana:syserr}. This fraction 
corresponds to 28\% in terms of the electon density, which is  
comparable to the statistical error in this region (see table~\ref{table:density}).
The steepest slope including this systematic effect is $-1.8$, and is still much
shallower than the NFW profile.

\cite{cavaliere-2009} analytically constructed 
a physical model of the ICM in hydrostatic equilibrium
with the dark matter potential,
in the context of dark-matter Jeans equilibria \citep{lapi-2009}
constrained by the {\it dark matter entropy} in a single power-law form
$K(r)\propto r^\alpha$,
as motivated by
N-body simulations of cold dark matter halos \citep{taylor-2001, faltenbacher-2007}.
Their {\it equilibrium} model
predicts a shallow gas density slope of
($d \ln n/d \ln r \sim -2.2$) at the entropy slope of $\sim1.1$ in a slow and constant accretion phase,
in good agreement with our observed gas density slope of $-2.29\pm0.18$
in the range of $r_{2500} \simlt r \simlt r_{500}$.
Beyond $r_{500}$ of Abell~1689, we calculated the ratio of the dark matter circular velocity 
to the gas sound velocity from the observed gas density and entropy slopes,
by following their model \citep[equation~7 in ][]{cavaliere-2009}.
The resulting ratio is $\sim1-2$,
which conflicts with our observations that the dark matter velocity is much
higher than the sound velocity. As we discuss in \S~\ref{section:dis},
the ICM in most regions at the outermost annulus
is likely {\it not} to be in hydrostatic equilibrium.
Therefore, some modifications to the model would be needed to describe our results outside $r_{500}$.


\subsubsection*{Entropy Profile}

The bottom panel in figure \ref{fig:prof} shows the entropy profile 
calculated as
\begin{equation}
K = \frac{kT}{n_{\rm e}^{2/3}}\; , 
\end{equation}
using the temperature and electron density profiles obtained above. 

The entropies in all directions
 increase from the center to the radius of $10\farcm0$.
The entropy distribution in the last radial bin is anisotropic, similarly to the temperature profiles.
The radial dependence of the Offset1 entropy obeys roughly $r^{1.1}$ 
(dashed line in figure~\ref{fig:prof}(c)), which is predicted by models 
of accretion shock heating \citep{tozzi-2001,ponman-2003}, while the entropies of 
the other regions (Offset2-Offset4) in the last radial bin are lower than those 
in their $6\farcm0$--$10\farcm0$ annuli.

\section{Discussion}
\label{section:dis}
\subsection{Thermodynamics in the Outskirts}
\label{subsec:ICM_outskirts}
In this section we discuss the distributions of gas temperature and 
entropy shown in figure \ref{fig:prof}, particularly focusing on their anisotropic distributions
in the last bin ($r_{500} \simlt r \simlt r_{\rm vir}$).

In the framework of the hierarchical clustering scenario based on CDM paradigm,
mass aggregation processes onto clusters, in particular along the large-scale
filamentary structures, are still on-going. 
Continuous mass accretion flows along filamentary structures, 
within which clusters are embedded, 
sometimes trigger {\it internal shocks} in gas within clusters.
It converts most of the tremendous amounts of kinetic energy into the
thermal energy to heat the ICM.
Therefore, the accretion flows are considered to play an important role in cluster evolution 
as well as in the ICM thermodynamics.
If the shock heating dominates in the cluster outskirts, 
the entropy generated in the shock  process becomes higher than those in the pre-shock regions. 
On the other hand, if the gas is adiabatically compressed 
during the infall with a sub-sonic velocity, 
the gas temperature increases but the entropy stays constant.

According to analytical models and simulations considering both the continuous accretion
 shock and adiabatic compression processes \citep[e.g.][]{tozzi-2001,voit-2002,borgani-2005},
the entropy profile is predicted to increase with $K\propto r^{1.1}$ by the accretion shock, 
as long as the initial entropy is low,
assuming that the ICM is in hydrostatic equilibrium and 
the kinetic energy of the infalling gas is instantly thermalized via shocks.
This radial dependence has been generally confirmed in observed profiles
 \cite[e.g. ][]{ponman-2003, pratt-2006} within $r_{500}$.
Interestingly, our entropy profile in the Offset1 direction coincides very well with the model 
prediction up to the virial radius.
It indicates a possibility that the ICM in the Offset1 $r_{500} \simlt r \simlt r_{{\rm vir}}$ region
is thermalized and close to the hydrostatic equilibrium.
Indeed, the outskirts temperature in the Offset1 direction agrees with the predictions of 
the analytical models \cite[e.g. ][]{komatsu-2001,tozzi-2001,ostriker-2005}
in which, under the hydrostatic equilibrium assumption, 
the temperature around the virial radius declines at most $50$ percent of the inner values.

Contrary to the Offset1, 
the sharp decline in both temperature and entropy in the Offset2-Offset4 regions 
indicates that the thermal pressure is not sufficient to balance the total gravity of the cluster.
If the gas is not completely thermalized, 
the gas temperature should be lower than the value needed to maintain hydrostatic equilibrium.
In this case, the accreted gas retains some fraction of the total energy in bulk motion,
and the bulk pressure partly supports the gravity. 
In fact, recent numerical simulations \cite[e.g.][]{vazza-2009} have shown 
that the kinetic energy 
of bulk motion carries $\sim 30\%$ of the total energy around the virial radius.
In addition to the bulk pressure, pressure of turbulent motions in the ICM may also contribute to 
balance the total gravity 
\cite[e.g. ][]{nagai-2007, piffaretti-2008, jeltema-2008, lau-2009,fang-2009,vazza-2009}.

\begin{figure}
  \begin{center}
   \includegraphics[width=0.55 \textwidth,angle=0,clip]{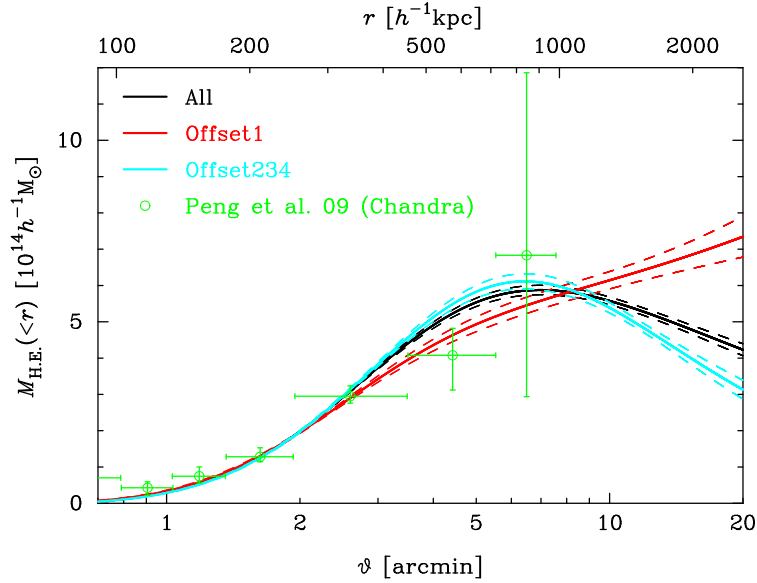} 
  \end{center}
  \caption{The hydrostatic equilibrium mass profile. The black, red and light blue colors represent the hydrostatic equilibrium mass for all, Offset1 and Offset234 regions, respectively. The solid and dashed lines indicate 
the best fit values and the 68\% CL uncertainty errors obtained from Monte Carlo simulations, respectively.
The green points are the model-independent mass based on the {\it Chandra} observation by \citet{peng-2009}, which agrees with our results. 
The {\it unphysical} declines in the cumulative masses (all and
 Offset234) indicate that the ICM of the outskirts is far from
 hydrostatic equilibrium we assumed. 
}
\label{fig:Mhe}
\end{figure}

In order to study the validity for the assumption of hydrostatic equilibrium, 
we measured the hydrostatic equilibrium masses (H.E. mass) for the all, Offset1 and Offset234 regions.
We here assume that 
the mass distribution is spherically symmetric and the ICM is in hydrostatic equilibrium at any radii.
The hydrostatic mass, $M_{\rm H.E.}$, is calculated from the parametric density and 
temperature profiles obtained by fitting data points in each azimuthal region (all, Offset1, \& Offset234) 
with models (Appendix \ref{appen:n+Tpara} 
and table \ref{table:nfit}), 
as follows
\begin{eqnarray}
 \frac{1}{\rho_g} \frac{d P_g}{d r} &=& - \frac{GM}{r^2} \label{eq:HE} \\
  P_{g} &=& \frac{\mu_e}{\mu} n_e k_B T, \nonumber
\end{eqnarray}
where $P_g$ is the gas pressure and $\rho_g=\mu_e m_p n_e$ is the gas mass density.
The errors (68\% CL uncertainty) are calculated with Monte Carlo simulations because model parameters are correlated with each other.
The resulting mass is shown in figure~\ref{fig:Mhe}.
The cumulative mass in the Offset1 direction increases with radius, 
while the ones in all and Offset234 regions {\it unphysically} decrease outside $\sim7\farcm0$--$8\farcm0$.
This {\it unphysical} decrease in the cumulative mass is due to the sharp temperature drops 
in the Offset234 region.
Here, in the parametric density model (equation \ref{eq:ne_pr}), we adopted a modified $\beta$ model
to reproduce the density flatness in outskirts ($r>10\farcm0$). The ratio of core radii
$r_{c,2}/r_{c,1}$ in equation \ref{eq:ne_pr} was fixed to be 3, since the flattening factor in which
$r_{c,2}$ is included was not constraind well owing to few data points.
However, even by choosing the ratio $\Delta(r_{c,2}/r_{c,1})=\pm1$,
the radii, at which the masses have maximum values, are changed only by $+7\%,
-3\%$ for Offset234 and $+12\%,-6\%$ for all regions, respectivly.
This is because the curvature of the parametric
density profile is insensitive to a choice of the ratio of core radii. 
Thus, the hydrostatic equilibrium we assumed is inadequate to describe the 
ICM in the outskirts except the Offset1 direction,
that is, most of the ICM in the cluster outskirts is far from hydrostatic equilibrium.

One alternative proposal to 
the bulk motion or turbulent motion as a cause of the non-hydrostatic
equilibrium would be convective instability in the cluster outskirts. 
We investigated the possibility of convective instability in the radial direction.
The convective motion in Offset234 is unstable outside $\sim9\farcm0$ 
but the time scale of the growing mode is comparable to 
the age of the universe
 at cluster redshift $z=0.1832$.
Therefore, the convective instability is negligible for the {\it unphysical} decrease in the cumulative mass.
Another possible proposal
is that ions have a higher temperature than that of electrons
\citep[]{fox-1997, ettori-1998, takizawa-1998}, 
and that the thermal pressure of ions supports the cluster mass. 
We computed the thermal equilibration time between electrons and ions through the Coulomb interaction.
In the cluster outskirts of Abell~1689, the time scale is given by
\begin{eqnarray}
t_{\rm ei}\sim 0.4 \left(\frac{n_e}{5\times10^{-5}~{\rm cm}^{-3}}\right)^{-1}  \left(\frac{k_B T}{2~{\rm keV}}\right)^{3/2}~{\rm Gyr} 
\end{eqnarray}
\cite[e.g. ][]{spitzer-1962,takizawa-1999,akahori-2009}.
The Coulomb interaction would provide us with the maximum time scale of thermal equilibrium, 
because there might be other processes, such as plasma instability, to facilitate the 
interaction between electrons and ions.
Since the resulting time is much shorter than the dynamical time scale, 
it is difficult to explain the {\it unphysical} decrease. 
Therefore, the bulk and/or turbulent motions would be the main source(s) of additional pressure
needed in the Offset234 directions.
In \S \ref{subsection:Pwl}, we shall discuss on this point again 
based on a joint X-ray and lensing analysis.

What makes the anisotropic distributions of gas temperature and entropy in the outskirts?
The internal shocks associated with mass accretion flows would heat the ICM. 
Numerical simulations \citep[e.g.][]{ryu-2003, molnar-2009} have shown that, 
the low-density gas in low-density regions,  so-called voids, 
accretes onto the vicinity of clusters by the gravitational pull, 
with an order of thousands of kilometers per second.
As a result, the internal shocks form 
outside virial radius at which gas density sharply changes.
It is also referred to as {\it virial shocks} \citep{ryu-2003}.
Meanwhile, the accretion flow along filamentary structures forms internal shocks
inside the virial radius.
These shocks migrate relatively deep into a cluster potential well,
because a large amount of matter, such as gas, galaxies and groups, accretes
from filaments \citep{ryu-2003}.
Therefore, the thermalization in the outskirts along the direction of filamentary
structure takes place faster.
Based on this accretion shock scenario,
the observed anisotropic distributions of gas temperature and entropy in the outskirts 
inside virial radius would be associated with large-scale structures.
Another possibility for the anisotropy is that infalling galaxies, especially 
massive galaxies, might be reservoirs of cold gas. 
The cold and dense gas, which is bound by the dark matter potential of galaxies, 
would decrease the emission-weighted temperature in the outskirts.
Although we excluded point sources in the spectral study, 
we cannot rule out this possibility only from the X-ray results. 
If this is the case, member galaxies around the virial radius ($10\farcm0$--$18\farcm0$)
in the Offset234 directions 
would be more numerous than those in the Offset1 direction.

\begin{figure*}
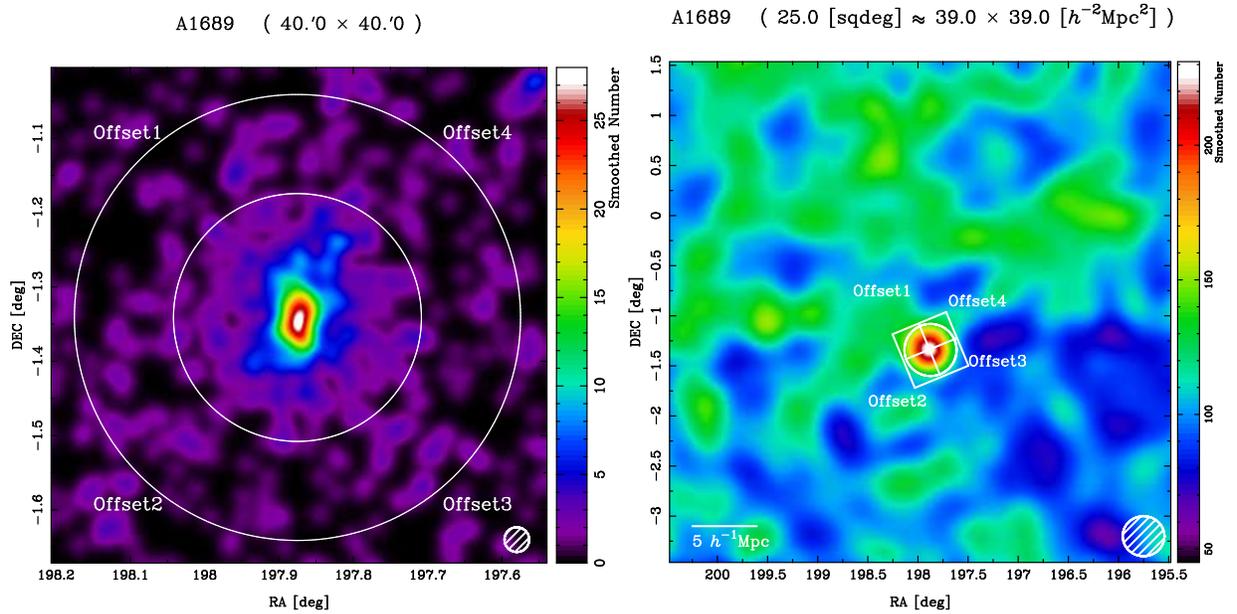

  \begin{center}
   \includegraphics[width=0.45 \textwidth,angle=0,clip]{./f9a.eps} 
   \includegraphics[width=0.44 \textwidth,angle=0,clip]{./f9b.eps} 
  \end{center}
  \caption{The two-dimensional density distributions 
of galaxies around A1689, retrieved from {\it SDSS} photometric data.
(left) : Distributions of red sequence galaxies of Abell~1689.
The box size is $40\farcm0\times40\farcm0$.
The annulus is the outskirts region for spectral fitting, 
where inner and outer circles are $10\farcm0$ and $18\farcm0$, respectively.
The central distribution is clearly elongated along the north-south direction.
We could not find a significant difference between projected distributions 
in the Offset1 and the other directions.
(right) : Distributions of galaxies whose photometric redshifts span in the range
 of $| z-z_{\rm c} | \leq 0.016$, where $z_{\rm c}$ is the cluster redshift.
The boxes represent the FOVs of XIS pointings.
The circle is the virial radius determined by lensing analysis.
A Gaussian smoothing scale is represented by the hatched circle at
 bottom right in each plot.
}
\label{fig:Mapsdss}
\end{figure*}

In order to search the large-scale structures or the anisotropic galaxy distribution
around the virial radius, we made maps of galaxy number density using 
the {\it SDSS} DR7 catalogue (Abazajian et al. 2009) from {\it SDSS} CasJobs
site\footnote{http://casjobs.sdss.org/}. 
We first looked into bright red-sequence galaxies which are expected to host cold gas.
We selected red-sequence galaxies by criteria of $|(g'-i') -
(-0.048i'+2.556)|<0.152$ and $i'<21~{\rm ABmag}$, 
where we used {\it psfMag} for magnitude 
and {\it modelcolor} for color.
We made projected galaxy distribution, smoothed with
a Gaussian kernel, $w=\exp(-r^2/r_g^2)$, of FWHM$=2(\ln2)^{1/2}r_g=1\farcm67$.
The resulting map is shown in the left panel of figure \ref{fig:Mapsdss}.
We also looked into the same map derived from {\it Subaru} data.
It is consistent with the {\it SDSS} map,
although a part of the outer region is not available in the {\it Subaru} data 
due to the limitation of its field-of-view.
In figure \ref{fig:Mapsdss} (left),
the central part of density map is clearly elongated in the north-south direction.
\citet{umetsu-2008} have found the same elongation in a two-dimensional 
mass distribution.
However, we could find neither an apparent feature 
nor a significant difference between the projected distributions 
in the Offset1 and the other directions in the cluster outskirts of $10\farcm0 < r < 18\farcm0$. 
Therefore, the temperature and entropy anisotropies in the outskirts are not
due to the cold gas in galaxies.

Next, we made a larger map to investigate the presence of large-scale structures associated
with Abell~1689.
\cite{lemze-2009} has shown that the maximum line-of-sight velocity of
infall galaxies is $\sim4000~{\rm km/s}$. 
Therefore, taking into account the uncertainty of redshift due
to line-of-sight velocities, we selected galaxies in the range of $|z-z_{\rm c}|
< \delta z =\sigma_{v,{\rm max}}(1+z_{\rm c})/c\simeq0.0158$, where $z_{\rm c}$ is the cluster
redshift, $z$ is a photometric redshift,
 $\sigma_{v,{\rm max}}=4000~{\rm km/s}$, and $c$ is the light velocity.
The mean photometric redshift for our sample is
$\bar{z}=0.18320\pm0.00004\pm0.00035$, where the first error is the
statistical error and the second one is the systematic error due to
photometric uncertainty of each galaxy.
Our sample of galaxies thus statistically represents a slice around Abell~1689 in redshift space.

The resulting map of galaxies smoothed with a Gaussian kernel of FWHM=20\farcm0 is
shown in the right panel of figure \ref{fig:Mapsdss}.
In the Offset1 direction, 
a filamentary overdensity region outside the virial radius is found, 
where the galaxy number density is more than 1.5 times higher than 
those in the other directions. 
In the Offset4 direction outside the FOV of XIS,
a narrow sheet of galaxies is also found, which is
connected to a northwest overdensity region. 
We note that the projected elongated direction of cluster mass and galaxies
(left panel of figure \ref{fig:Mapsdss}) does not coincide with the
filamentary direction (right panel), suggesting that a mass structure
seen in the central region of a cluster is not necessarily connected
directly to a filament. 
Our result does not change even when we choose twice or half $\delta z$
in the galaxy selection.
We also tried to investigate the line-of-sight filamentary structure 
by a Monte -Carlo simulation computing the redshift distribution of galaxies 
with photometric errors. 
However, we could not obtain reliable results within a redshift
resolution smaller than $\sim10~{\rm Mpc}$.
We could not see the line-of-sight structure from the spectral data of {\it SDSS}
either, because available number of galaxies is too small. 

The high temperature and entropy region in the outskirts 
is clearly correlated with the galaxy overdensity region associated with
the large-scale structure outside Abell~1689, 
as shown in figure~\ref{fig:Map_kt_sdss}.
This indicates that the ICM in the outskirts is significantly
affected by surrounding environments of galaxy clusters, such as
the filamentary structures and the low density void regions.
The large-scale structure would play an important role in 
the thermalization process of the ICM in the outskirts.
In particular, our result suggests that the thermalization in the outskirts along the
filamentary structure takes place faster than that in the void region.

\subsection{A Joint X-ray and Lensing Analysis}
\label{sec:joint}

In this subsection, we carry out a joint X-ray and lensing analysis, incorporating
{\it Subaru/Suprime-Cam} and {\it HST/ACS} data. 
Abell~1689 has been the focus of intensive lensing studies in recent years
\citep[e.g,
][]{broadhurst-2005a,broadhurst-2005b,limousin-2007,umetsu-2008,corless-2009}.
It has been shown by 
\cite{broadhurst-2005b} and \cite{umetsu-2008} 
that joint lensing profiles of Abell~1689 obtained from their ACS and Subaru data
with sufficient quality are
consistent with a continuously steepening density profile
over a wide range of radii, $r=10-2000$\,kpc$\,h^{-1}$,
well described by the general NFW profile
\citep{navrro-1996,navrro-1997}, whereas the 
singular isothermal sphere model is strongly disfavored.
They also revealed that the concentration parameter of the NFW
profile, namely the ratio of the virial radius to the scale radius, 
$c_{\rm vir}=r_{\rm vir}/r_s$,
is much higher than predicted by cosmological $N$-body simulations 
\cite[e.g., ][]{bullock-2001,neto-2007}. 
We note that the virial radius, $r_{\rm vir}$, within which the mean interior
density is $\Delta_{\rm vir}\simeq110$ \citep{nakamura-1997}
times the critical mass density, $\rho_{\rm cr} (z)$,
at a cluster redshift is given by
\begin{eqnarray}
M_{\rm vir}= \frac{4}{3} \pi \Delta_{\rm vir} \rho_{\rm cr}(z) r_{\rm
 vir}^3. \label{eq:viral}
\end{eqnarray}
In the present paper, the full lensing constraints derived from the joint ACS and Subaru
data allow us to compare, for the first time,
X-ray observations with the 
total mass profile for the entire cluster,
from the cluster center to the virial radius.
The lensing analysis is 
free from any assumptions about the dynamical state of the cluster,
so that a joint X-ray and lensing analysis provides a powerful
diagnostic of the ICM state and any systematic offsets between the two
mass determination methods.

Here we first summarize our lensing work on Abell~1689 before presenting the
results from our joint analysis.
\cite{umetsu-2008} combined HST/ACS strong lensing data with 
Subaru weak lensing distortion and magnification data in a two-dimensional
analysis to reconstruct the projected mass profile.
Their full lensing method, assuming the spherical symmetry, yields best-fit
NFW model parameters of 
$M_{\rm vir}=1.47^{+0.59}_{-0.33} \times 10^{15} M_\odot\,h^{-1}$ and
$c_{\rm vir}=12.7\pm 2.9$ (including both statistical and systematic
uncertainties; 
see also \cite{lemze-2009}), which properly reproduce the observed
Einstein radius of $\theta_{\rm E}=45\arcsec$ for $z_s=1$
\citep{broadhurst-2005a}.
Here we deproject the two-dimensional mass profile of \cite{umetsu-2008}
and obtain a non-parametric $M_{\rm 3D}$ profile
simply assuming spherical symmetry \citep{broadhurst-2008,umetsu-2009b}.
This method is based on the fact that
the surface-mass density $\Sigma_m(R)$ is related to the
three-dimensional mass density $\rho(r)$ by an Abel integral transform;
or equivalently, one finds that
the three-dimensional mass $M_{\rm 3D}(<r)$
out to spherical radius $r$ is written in terms of $\Sigma_m(R)$ as
\begin{equation}
\label{eq:m3d}
M_{\rm 3D}(<r) = 2\pi\int_0^r\! dRR\Sigma_m(R)
-4\int_r^\infty\!dRR f
\left(
\frac{R}{r}
\right)
\Sigma_m(R),
\end{equation}
where
$f(x)=(x^2-1)^{-1/2}-\tan^{-1}(x^2-1)^{-1/2}$
\citep{broadhurst-2008}.
We propagate errors on $\Sigma_m(R)$ using
Monte-Carlo techniques taking into account
the error covariance matrix of the full lensing constraints.
This deprojection method allows us to derive in a non-parametric way
three-dimensional virial quantities ($r_{\rm vir}, M_{\rm vir}$)
and values of $M_{\Delta}=M_{\rm 3D}(<r_\Delta)$
within a sphere of a fixed mean interior overdensity $\Delta$
with
respect to the critical density of the universe at the cluster
redshift. 
The resulting parameters of 
the non-parametric spherical, as well as NFW, models
are summarized in table~\ref{table:lensmass}.

\subsubsection*{Mass Comparison}

Here we compare our X-ray hydrostatic mass measurements
(in all regions; \S \ref{subsec:ICM_outskirts})  
with the spherical-lensing and NFW mass profiles
in figure \ref{fig:Mhe_vs_Mwl}.
As discussed in \S \ref{subsec:ICM_outskirts}, the hydrostatic
 equilibrium assumption 
 is invalid in the cluster outskirts and hence,
a mass comparison is limited within the radius $\sim7\arcmin$,  
inside which the hydrostatic mass increases with radius.
The hydrostatic mass is systematically lower than the lensing mass at
 all radii. 
In particular, the hydrostatic mass in the central region ($r\simlt
 2\farcm0$) 
 is significantly lower than the lensing one ($M_{\rm H.E.}/M_{\rm
 lens}\simlt60\%$).   
Since our hydrostatic mass estimates 
agree with the latest {\it Chandra} results
 \citep{peng-2009} in the overlapping region, 
this would not be due to the low angular resolution of the 
{\it Suzaku} satellite. 
At intermediate radii of $3\farcm0\simlt r \simlt 7\farcm0$, 
the hydrostatic to lensing mass ratio, $M_{\rm H.E.}/M_{\rm lens}$,
is in a range of $\sim 60-90\%$.   
Some numerical simulations
\cite[e.g. ][]{nagai-2007,piffaretti-2008,fang-2009} have shown that
X-ray hydrostatic-equilibrium mass estimates 
would underestimate the actual
cluster mass,
because the turbulent pressure and the bulk/rotational kinetic
pressure  partially support the gravity of the cluster.
For instance, 
\cite{nagai-2007} found that the hydrostatic mass is
biased low, and is underestimated on average 
by $0.121\pm0.136$ and $0.163\pm0.095$ at overdensities $\Delta=2500$ and
$500$, respectively,
for a sample of 12 simulated clusters at redshift $z=0$.
\cite{piffaretti-2008} found that the hydrostatic mass derived 
from the $\beta$-model gas density profile
 and spectroscopic-like temperature model, 
similarly to our modeling,
is biased low by 
$12\pm11$\% and $27\pm11$\% at overdensities of $\Delta=2500$
and $500$, respectively. 
They found that,
on average, the systematic underestimate bias of $M_{\rm H.E.}$
(with respect to the actual cluster mass)
increases with decreasing $\Delta$, or increasing the pivot radius
$r_\Delta$.

We find that 
the degree of systematic bias 
in the X-ray mass estimate, $M_{\rm H.E.}/M_{\rm
lens}$, is in rough agreement with the 
results from numerical simulations,  
except for the central region.
However, such a comparison with simulated clusters in the central region
would be practically difficult owing to {\it over-cooling} problems in
cosmological hydrodynamical simulations, failing to reproduce the
observed ICM features.
If the ICM is not in thermal balance with the gravity
and an additional pressure support
is provided by the gas bulk/turbulent motions,
as suggested by simulations, 
the X-ray micro-calorimeter instrument (SXS)
on the next Japanese X-ray satellite {\it  ASTRO-H} 
\citep{takahashi-2008}
will directly detect them
and 
measure their contributions to the pressure balance.

Another possible cause of the mass discrepancy 
is the triaxial structure of clusters.
If the major axis of a triaxial halo is pointed to the observer,
then this orientation boosts the projected surface mass density
$\Sigma_m(\theta)$
and hence the lensing signal. As a result, 
halo triaxiality and orientation bias can 
affect the lensing-based mass and concentration measurements \citep[e.g, ][]{oguri-2005}.
\cite{oguri-2009} demonstrated in the context of the $\Lambda$CDM model
that clusters with large Einstein radii ($\theta_{\rm E}\ge 20\arcsec$),
or the so-called {\it superlenses}, are likely to be highly biased, with major
axes preferentially aligned with the line-of-sight. 
The level of correction for the orientation bias 
in lensing mass estimates
(obtained a priori assuming the spherical NFW model)
is derived from a semi-analytical representation of simulated CDM
triaxial halos by \cite{oguri-2005}.
\cite{oguri-2005} fitted the two-dimensional mass map 
of Umetsu \& Broadhurst (2008) with a triaxial halo model and obtained
$M_{\rm vir}=1.26_{-0.49}^{+0.28}\times 10^{15}h^{-1}M_{\odot}$ and $c_{\rm vir} =
16.9^{+2.2}_{-12.9}$ (table~\ref{table:lensmass})
where the quoted errors here are statistical only (68.3\% CL).
This triaxial description of the full lensing constraints presented in
Umetsu \& Broadhurst (2008) is in better agreement with our X-ray
results.
On the other hand, more recently,
\cite{peng-2009} demonstrated that a simple triaxial modeling of 
{\it Chandra} X-ray data can relax the mass discrepancy with respect to
strong and weak lensing.
They found that the total spherical mass and the spherically-averaged
mass density are essentially unchanged under different triaxiality
assumptions, as previously reported by \cite{piffaretti-2003} and \cite{gavazzi-2005}.
By assuming a prolate symmetry of
the gas structure with axis ratio $0.7$,
they found that 
their X-ray mass estimate agrees with those
of strong and weak lensing within 1\% ($-1\sigma$) 
and 25\% ($+1\sigma$), respectively.

Using the triaxial halo model of \cite{oguri-2005}, 
the X-ray to lensing mass ratio becomes consistent with
unity at larger radii within the errors;  
however, it is still much lower than unity in the central region due to the inferred high
concentration parameter.
Based on independent weak lensing data,
\cite{corless-2009} also obtained high concentrations from their triaxial halo
modeling in conjunction with the strong lensing prior on the Einstein
radius.
Due to the tight strong lensing constraints that prefer a
highly-concentrated mass distribution, it is difficult to
explain the mass discrepancy in the central region.

\subsubsection*{Gas Fraction}

\begin{figure*}
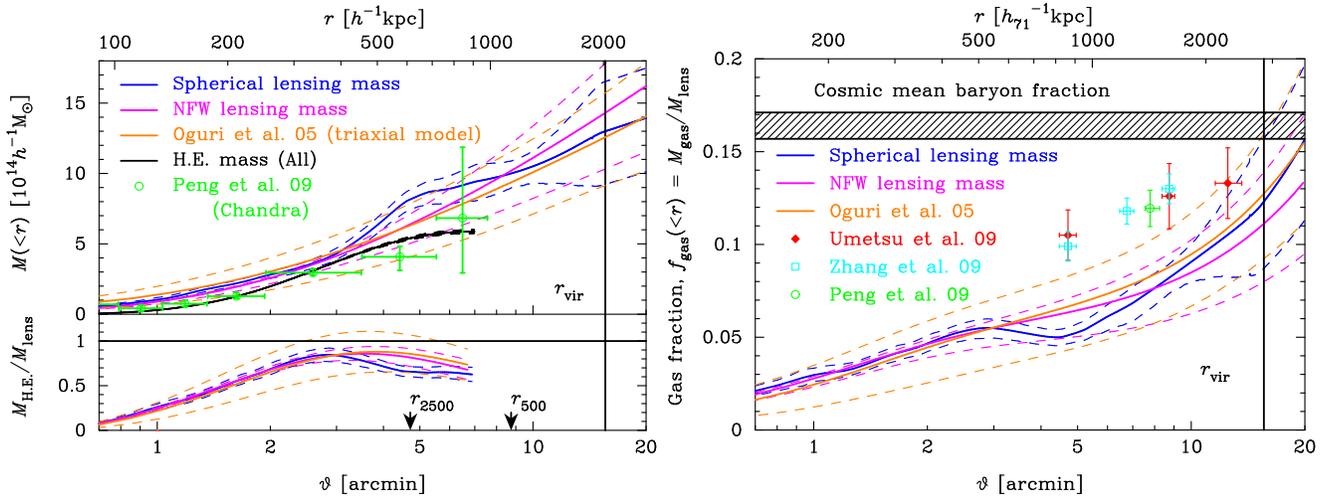

  \begin{center}
   \includegraphics[width=0.48 \textwidth,angle=0,clip]{./f11a.eps} 
   \includegraphics[width=0.48 \textwidth,angle=0,clip]{./f11b.eps} 
  \end{center}
  \caption{  (left): Comparison of lensing mass, obtained from joint strong and weak 
lensing analysis, and hydrostatic mass. Blue, magenta and orange colors represent deprojected 
spherical lensing mass, NFW one, and triaxial halo model \citep{oguri-2005}, respectively.
Black line represents the hydrostatic mass in all directions within 
the radius $r\sim7\farcm0$, outside which the hydrostatic equilibrium is not adequate 
any more (figure \ref{fig:Mhe}). 
The solid and dashed lines are the best fit values and the 68\% CL uncertainty 
errors obtained from Monte Carlo simulations, respectively.
Green points are the hydrostatic mass using {\it Chandra} data \citep{peng-2009}.
The arrows denote the overdensity radii $r_{2500}$ and $r_{500}$ determined by 
deprojected spherical lensing mass. 
 (right): Cumulative gas mass fraction. Blue, magenta and orange colors represent gas mass 
fractions using deprojected spherical lensing mass, NFW one, and triaxial halo 
model \citep{oguri-2005},  respectively. 
The solid and dashed lines are the same as left panel.
Red points are the gas mass fraction based on {\it AMiBA} SZE measurement 
and {\it Subaru} lensing analysis for 4 clusters  \citep{umetsu-2009a}, 
at three overdensities $\Delta=2500, 500~\&~200$.
Light blue points are the gas mass fraction based on {\it XMM-Newton} X-ray 
and {\it Subaru} lensing analysis for 12 clusters of {\it LoCuSS} project \citep{zhang-2010}, 
at three overdensities $\Delta=2500, 1000~\&~500$.
Green point denotes the gas fraction using {\it Chandra} hydrostatic mass 
for A1689 \citep{peng-2009}. 
In this figure, we use $h_{71} = H_0 / 71\; \rm{km\; s^{-1}\; Mpc^{-1}}=1.0$.
}
\label{fig:Mhe_vs_Mwl}
\end{figure*}

From our combined X-ray and lensing data set,
we measure the cumulative gas mass fraction,  
\begin{eqnarray}
 f_{\rm gas} (<r) = \frac{M_{\rm gas}(<r)}{M_{\rm lens}(<r)}, 
\end{eqnarray} 
where $M_{\rm gas}(<)$ and $M_{\rm lens}(<r)$ are the gas and lensing
masses contained within a sphere of radius $r$.
The gas mass is calculated by fitting a modified $\beta$ model to
the electron number density profile (Appendix \ref{appen:n+Tpara}). 
The errors are obtained by 1000 Monte-Carlo simulations taking into
account the error covariance matrix.
The resulting gas fraction, shown in the right panel of figure~\ref{fig:Mhe_vs_Mwl},
increases toward the outskirts, but does not reach the cosmic mean
baryon fraction \citep[WMAP5;][]{komatsu-2009} within the virial radius.
In figure~\ref{fig:Mhe_vs_Mwl} (right), 
we also compare our result with those from previous
statistical studies
\citep{zhang-2010,umetsu-2009a}:
the mean gas fractions for 12 clusters from the Local Substructure
Survey (LoCuSS) 
based on a joint {\it XMM-Newton} X-ray and {\it Subaru/Suprime-Cam}
weak lensing analysis \citep{zhang-2010}
and those for 4 high-mass clusters ($M_{\rm vir}\simgt
10^{15}M_\odot\,h^{-1}$) 
from a joint Sunyaev- Z'eldovich effect (SZE) {\it AMiBA} and {\it Subaru} 
lensing analysis \citep{umetsu-2009a}. 
We note that these gas fraction measurements, based on the
lensing-derived total mass, are 
independent of the cluster dynamical state.
The mean gas fractions from the two statistical studies are in close
agreement with each other. 
On the other hand, our gas fraction measurements of Abell~1689 
based on the joint X-ray/lensing analysis,
assuming spherical symmetry,
are about half of the mean cluster values
derived from \cite{zhang-2010} and \cite{umetsu-2009a}.
For the case of 
the triaxial mass model \citep{oguri-2005},
the upper bounds of the gas fraction come closer to,
and are marginally consistent with,
the mean values derived by statistical work.
Recently
\cite{bode-2009} proposed a method of modeling the 
ICM distributions in gravitational potential
wells of clusters, including 
star formation, energy input, and nonthermal pressure support
calibrated by numerical simulations and 
X-ray observations. In their standard model, the gas 
initially contained inside the virial radius expands outward. 
Consequently, the cluster baryon fraction reaches the cosmic mean
at $r\approx 1.2 r_{\rm vir}$, which is in rough agreement with our
result.

\subsubsection*{Pressure and Entropy Profiles Derived from Gravitational
   Lensing}
\label{subsection:Pwl}

We compare gas thermal pressure obtained from the two X-ray observables, temperature and 
electron number density, with a pressure calculated from the lensing mass profile 
assuming the hydrostatic equilibrium (hereafter lensing pressure). 
The lensing pressure is a total pressure to balance fully 
the gravity of the cluster lensing mass \citep[see also][]{molnar-2010}.
We calculated the lensing pressure profile, $P_{\rm lens}(r)$, by integrating eq. (\ref{eq:HE}) 
with a boundary condition that 
the external pressure is zero at $r=40\farcm0$ ($\approx 2\theta_{\rm vir}$).
The inner pressure is little affected by choices of the outer
boundary condition as long as the pressure outside the virial radius is
significantly lower than the values at inner radii.
We extrapolated the observed lensing mass and density profile outside the field-of-view.
The errors were computed by 1000 Monte Carlo simulations.  

\begin{figure*}
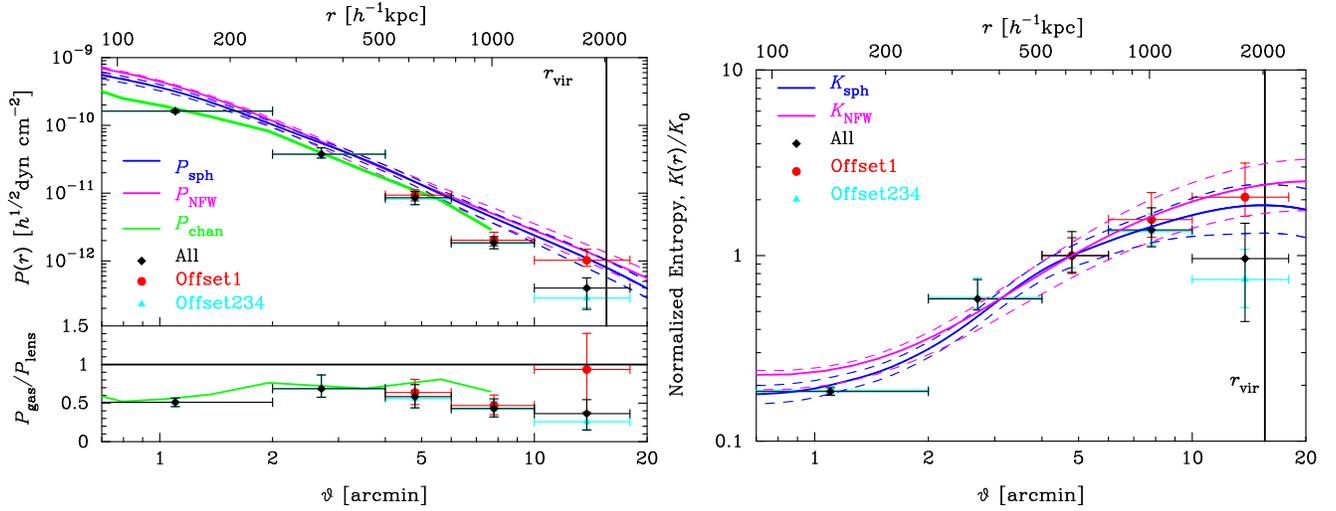

  \begin{center}
   \includegraphics[width=0.48 \textwidth,angle=0,clip]{./f12a.eps} 
   \includegraphics[width=0.48 \textwidth,angle=0,clip]{./f12b.eps} 
  \end{center}
\caption{ (left): Comparison of the gas thermal pressure 
with the total pressure expected from lensing masses with hydrostatic equilibrium assumption.
Blue and magenta colors are the pressure calculated from deprojected spherical lensing mass and NFW one, respectively.
Solid and dashed lines represent the best and 
$\pm1\sigma$
errors, respectively.
Black, red and light blue points are 
observed thermal pressures (68\% confidence errors)
in the all, Offset1 and Offset234, respectively.
Green line in the left panel is the pressure of {\it Chandra} data \citep{peng-2009}.
(right): The normalized entropy profiles. The colors are the same as those in the left panel.
The slope of the entropy in Offset1 is in good agreement with 
that expected from lensing masses with hydrostatic equilibrium assumption.
}
\label{fig:PTS}
\end{figure*}

The lensing pressure is systematically higher than the gas thermal pressure at any radii
 (the left panel of figure \ref{fig:PTS}).
Within the radius of $r_{2500}$, the gas thermal pressure is 
at most $50$ -- $60\%$ of the lensing pressure. 
In the outskirts around the virial radius,
the gas pressure in Offset1 is comparable to lensing one,
while ones in the Offset234 and all regions is $\sim30$ -- $40\%$.
This is inconsistent with a result from SZE analysis of clusters;
a stacked SZE signal profile 
for galaxy clusters using WMAP three-year data
\citep{afshordi-2007, atrio-Barandela-2008} has shown 
that the gas closely follows the dark matter distribution out to the virial
radius.

If we assume that the mass discrepancy is fully explained
by the gradient of additional pressure supports such as turbulent and/or bulk pressures,
they contribute up to
$\sim40$ -- $50\%$ of the lensing pressures 
within $r_{500}$ and $\sim60$ -- $70\%$ around the virial radius, respectively.
Hydrodynamic numerical simulation \citep{lau-2009}
has shown the fraction of turbulent pressure to total pressure on average 
increases up to $\sim30\%$ at $2r_{500}$, roughly corresponding to $r_{\rm vir}$.
They also found that the ratio at $r_{500}$ spans in 
a range from $6\%$ to $15\%$ for simulated relaxed clusters 
and from $9\%$ to $24\%$ for unrelaxed clusters, respectively.
\cite{dolag-2005} has found that the turbulent energy accounts for  $\sim5$ -- $30\%$ 
of the thermal energy content.
\cite{vazza-2009} has shown that the bulk energy changes 
from $\sim10\%$ of the total energy at the center to $\sim30\%$ at the virial radius
and the turbulent energy ($\simlt 5\%$) is less than the kinetic one.
The fractions of additional non-thermal pressure we derived above 
are slightly higher than
those in simulated clusters. 
They should be considered as upper limits, 
taking into consideration the possibility of triaxial halo model, 
that decreases the amount of non-thermal pressure relative to total lensing one.

We next computed entropy profiles from the lensing mass profile. 
Since the fraction of thermal pressure to the total one has not yet been revealed in reality,
we cannot calculate the entropy of the thermal gas using lensing mass.
We therefore compared the normalized gas entropy with lensing entropy ($K_{\rm lens}(r)$),
as shown in the right panel of figure \ref{fig:PTS}.
The normalization is applied at the third radial bin.
The lensing entropy surprisingly agrees with the normalized
 gas entropy in the Offset1 direction out to the virial radius. 
The normalized entropies in the Offset234 and all regions also trace the lensing entropy
up to $\sim r_{500}$, while, in the outskirts, observed entropies deviate downward.
Although we cannot constrain from figure \ref{fig:PTS}
 how much thermal energy is converted from the shock energy,
when we remember that the slope of gas entropy in Offset1 also coincides with 
the accretion shock models \citep[][\S \ref{subsection:ana:prof}]{tozzi-2001,voit-2002,borgani-2005},
our result indicates that the entropy generation by accretion shocks would occur, tracing the 
cluster mass profile.

\section{Summary}
\label{section:summary}
We observed Abell~1689 with \emph{Suzaku} and detected the ICM emission
out to the virial radius. Then, 
we derived radial profiles of temperature, electron density, and entropy, 
and conducted a joint X-ray and lensing analysis.
Our conclusions in this work can be summarized as below. 
\begin{itemize}
\item The ICM emission out to the virial radius was detected with $4.0 \sigma$ significance.
\item The temperature gradually decreases toward the virial radius from $\sim 10$ keV to $\sim 2$ keV.
\item The temperature variations in azimuthal direction becomes significant in $r_{500} \simlt r \simlt r_{\rm vir}$:
       the temperatures of Offset1 ($\sim5.4~$keV) and Offset234 ($\sim1.7~$keV) 
       are about half and one-sixth of $10.4$ keV which is an emission-weighted 
       temperature of $2\farcm0$--$4\farcm0$ and $4\farcm0$--$6\farcm0$ regions, respectively.
\item The temperature profile in the Offset1 direction is in good agreement with the scaled 
       temperature one by \cite{pratt-2007}, while the temperatures in the other directions
       (Offset2 - Offset4) around the virial radius are lower than the scaled value by a factor of $\sim 2$. 
\item The electron density profile decreases down to 
${\mathcal O}(10^{-5})~{\rm cm^{-3}}$  around the virial radius.
\item The electron densities in the four offset directions are consistent 
      with one another within errors in any circular annulus.
\item The slope of the electron number density is $-2.29\pm0.18$ in
      intermediate range of $r_{2500} \simlt r \simlt r_{500}$.
      In the outer annulus of $r_{500} \simlt r \simlt r_{\rm vir}$, the density slope, 
      ${-1.24_{-0.56}^{+0.23}}$ where the lower limit is the
      systematic error due to the contamination of X-ray emissions from
      point sources and inner region,  becomes shallower. 
      This outskirts value is significantly 
      shallower than the universal matter density profile $\rho\propto r^{-3}$ by 
      Navarro, Frenk, \& White (1996; 1997).
      
\item The entropies in all directions
 increase from the center to the radius of $10\farcm0$.
The entropy distribution in the last radial bin
    ($10\farcm0$--$18\farcm0$) is anisotropic, similarly to the
     temperature profiles.

\item The entropy profile of the Offset1 direction increases as $r^{1.1}$, as predicted by models 
      of accretion shock heating \citep{tozzi-2001,ponman-2003}, while the entropies of
      Offset2-Offset4 in the outskirts are lower than those in their annuli 
      of $6\farcm0$--$10\farcm0$.
      The slopes of the entropy profiles in the Offset1 direction and the other directions 
      match those predicted by lensing spherical masses with the
      hydrostatic equilibrium assumption, 
      from the core to the virial radius and to $\sim r_{500}$, respectively.
\item The cumulative hydrostatic mass in the Offset1 direction increases with radius, 
      while the ones in the other (Offset234) regions {\it unphysically} decrease outside 
     $\sim9\farcm0$. The ICM in the cluster outskirts, except the Offset1 direction,
     is far from hydrostatic equilibrium.    
\item The anisotropic distributions of gas temperature and entropy in the outskirts 
      are clearly correlated with the large-scale structure associated
      with Abell~1689.  In the Offset1 (northeast) direction, the hot and high entropy 
      region around the virial radius
      is connected to the overdensity region of galaxies in the 
      {\it SDSS} map,
      whereas the low-density region (void) outside the virial radius is found
      in the other directions.
      It indicates that the thermalization in the outskirts along the
      filamentary structure takes place faster than that in the void region.

\item The hydrostatic to spherical lensing mass ratio, $M_{\rm H.E.}/M_{\rm lens}$, 
      at intermediate radii of $3\farcm0\simlt r \simlt 7\farcm0$, 
      is in a range of $\sim 60$ -- $90\%$.  
      Using the triaxial halo model, the X-ray to lensing mass ratio in the same annulus 
      is consistent with unity within errors.
      To the contrary, the ratio within $2\farcm0$,  
      $M_{\rm H.E.}/M_{\rm lens}\simlt60\%$,  
      is significantly biased low, and the discrepancy cannot be compromised   
      by either the spherical or triaxial models

\item  The gas fraction computed from combined gas density and spherical lensing mass 
       does not reach the cosmic mean baryon fraction within the virial radius. 
       Although our fraction is lower than the mean value of previous statistical studies 
       based on lensing mass, the upper bounds of the gas fraction derived from 
       the triaxial mass model \citep{oguri-2005}
       come closer to, and are marginally consistent with, the statistical works.

\item The thermal gas pressure within $\sim r_{500}$ 
      is at most $\sim 40$ -- $60\%$ of the total pressure predicted by spherical lensing mass model.
      In the cluster outskirts outside $r_{500}$, 
      the thermal pressure in the Offset1 direction is comparable to total pressure, 
   whereas it is $\sim 30$ -- $40\%$ of the total pressure in the other directions.
      Taking into account the triaxial model, 
 these values give the lower limits.
 With such values the thermal pressure would be
 insufficient to balance the total gravity of the cluster, requiring additional pressure support(s).

\end{itemize}

\acknowledgments

We are very grateful to Molnar, S., M. for helpful comments. 
MK acknowledges support from a grant based on the Special Postdoctoral 
Researchers Program of RIKEN. 
The present work is supported in part
by the Grant-in-Aid for Scientific Research (S), No. 18104004.
NO and MT are in part supported by a Grant-in-Aid from the
  Ministry of Education, Culture, Sports, Science, and Technology of
  Japan (NO: 20740099; MT: 19740096). 
KU acknowledges support by the National Science Council of Taiwan
under the grant NSC95-2112-M-001-074-MY2.
This work is supported by a Grant-in-Aid for the COE Program
``Exploring New Science by Bridging Particle-Matter Hierarchy'' and
G-COE Program ``Weaving Science Web beyond Particle-Matter Hierarchy''
in Tohoku University, funded by the Ministry of Education, Science,
Sports and Culture of Japan.  This work is in part supported by a
Grant-in-Aid for Science Research in a Priority Area "Probing the Dark
Energy through an Extremely Wide and Deep Survey with \emph{Subaru}
Telescope" (18072001) from the Ministry of Education, Culture, Sports,
Science, and Technology of Japan.

\appendix
\section{Simulation of surface brightness profile
\label{appen:sim-sb}}
We simulated ICM emissions from Abell~1689 on XIS-FI (XIS0 + XIS3)
with \verb+xissim+, which spatially follow the double-beta model determined 
from the 0.5--10 keV {\it XMM-Newton} MOS1 image (\S~\ref{subsection:obs:reduction}).
As a spectrum model, we used 6.2 keV {\it apec} model (metal abundance is 0.3 solar)
with a flux of $2.5 \times 10^{-11}$ ergs cm$^{2}$ s$^{-1}$ (integrated within $30'$ 
in radius), which was determined from simultaneous fit to XIS-FI (averaged over XIS0 and XIS3) 
and XIS1 spectra of the Offset1 observation, extracted from whole XIS regions. 
Since Offset1 observation does not cover the central $\sim 4'$ region, the spectrum
model reproduces the emission from cluster outskirts. 

\begin{figure}
  \begin{center}
   \includegraphics[width=0.37\textwidth,angle=0,clip]{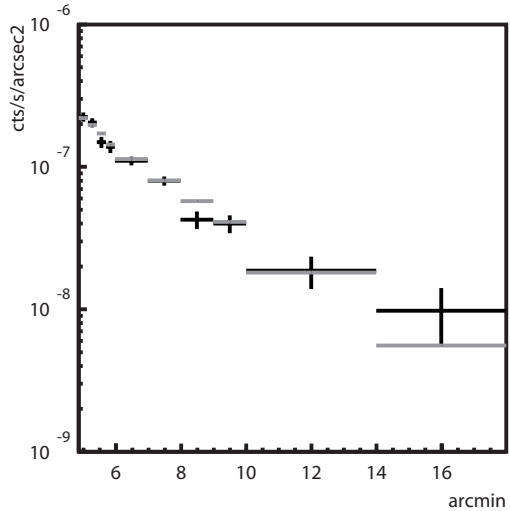} 
  \end{center}
  \caption{
Simulated 6.2 keV {\it apec} (gray) profile is shown, which follows the 
double-beta profile determined from the 0.5--10 keV {\it XMM-Newton} MOS1 
image as described in \S~\ref{subsection:obs:reduction}.
For comparison,
background-subtracted 0.5--10 keV surface brightness profile of
XIS-FI averaged over XIS0 and XIS3 (black) is also shown. 
In this black profile, the residual contribution of point sources estimated
in \S~\ref{subsection:ana:syserr} is also subtracted. 
In the error bars of the red profile, 1$\sigma$ uncertainties of the XRB and NXB
are added in quadrature to the corresponding statistical 1$\sigma$  errors. 
}
\label{fig:prof_sim}
\end{figure}

As shown in figure~\ref{fig:prof_sim}, the simulated profile is in 
good agreement with the observation. 
This indicates that the observed signals
around the virial radius can be described by extrapolating the double-beta profile. 
From another simulation which used the same double-beta model but truncated at 
$10'$, we found that the stray signals which originate from $< 10'$ regions 
contribute only 22\% of signals in $> 10'$ regions. 

In order to study an effect of temperature gradient expected in the ICM, we
also simulated XIS-FI (XIS0 + XIS3) profiles of 9 keV and 2 keV thermal emission 
with {\it apec} model (metal abundance is 0.3 solar). When normalization of these 
two simulated profiles are scaled to match at the center, the 9 keV profile
is higher than the 2 keV one by 10\% outside $10'$. Therefore, the temperature
gradient does not affect the simulated profile significantly.

\section{Simulation of point sources
\label{appen:sim-pointsrc}}
Out of the 49 point sources in table~\ref{table:pntsrc}, 
35 sources have counterparts in the Incremental Second XMM-Newton 
serendipitous survey catalog \citep[2XMMi,][]{watson-2009}, on which flux for 5 energy bands, 
0.2--0.5 keV, 0.5--1.0 keV, 1.0--2.0 keV, 2.0--4.5 keV,and 4.5--12.0 keV, are available from 
a HEASARC website\footnote{http://heasarc.gsfc.nasa.gov/docs/archive.html}.
For sources which are not on the 2XMMia catalog, we estimated their 0.2--2.0 keV flux by 
comparing their background-subtracted counts in the circular $1'$ regions in the PSPC 
total band (0.1--2.0 keV) image and that of source 7 (table~\ref{table:pntsrc}) of which flux 
is known. From the estimated 0.2--2.0 keV flux, we further estimated flux for the 5 energy bands,
assuming a power law spectrum with  photon index of 1.4. 

\begin{figure}
  \begin{center}
   \includegraphics[width=0.37\textwidth,angle=0,clip]{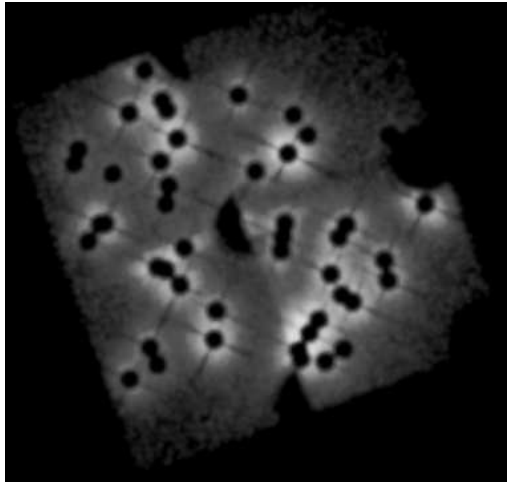} 
  \end{center}
  \caption{
A simulated 0.5--10 keV XIS-FI (XIS0 $+$ XIS3) mosaic image of point sources,
smoothed with a gaussian kernel of $\sigma = 10''$. 
$1'$ circular regions around these sources are excluded. Simulated exposures of the
4 observations are set 10 times longer than the actual ones. 
}
\label{fig:pntsrc_sim}
\end{figure}

\begin{figure}
  \begin{center}
   \includegraphics[width=0.23 \textwidth,angle=0,clip]{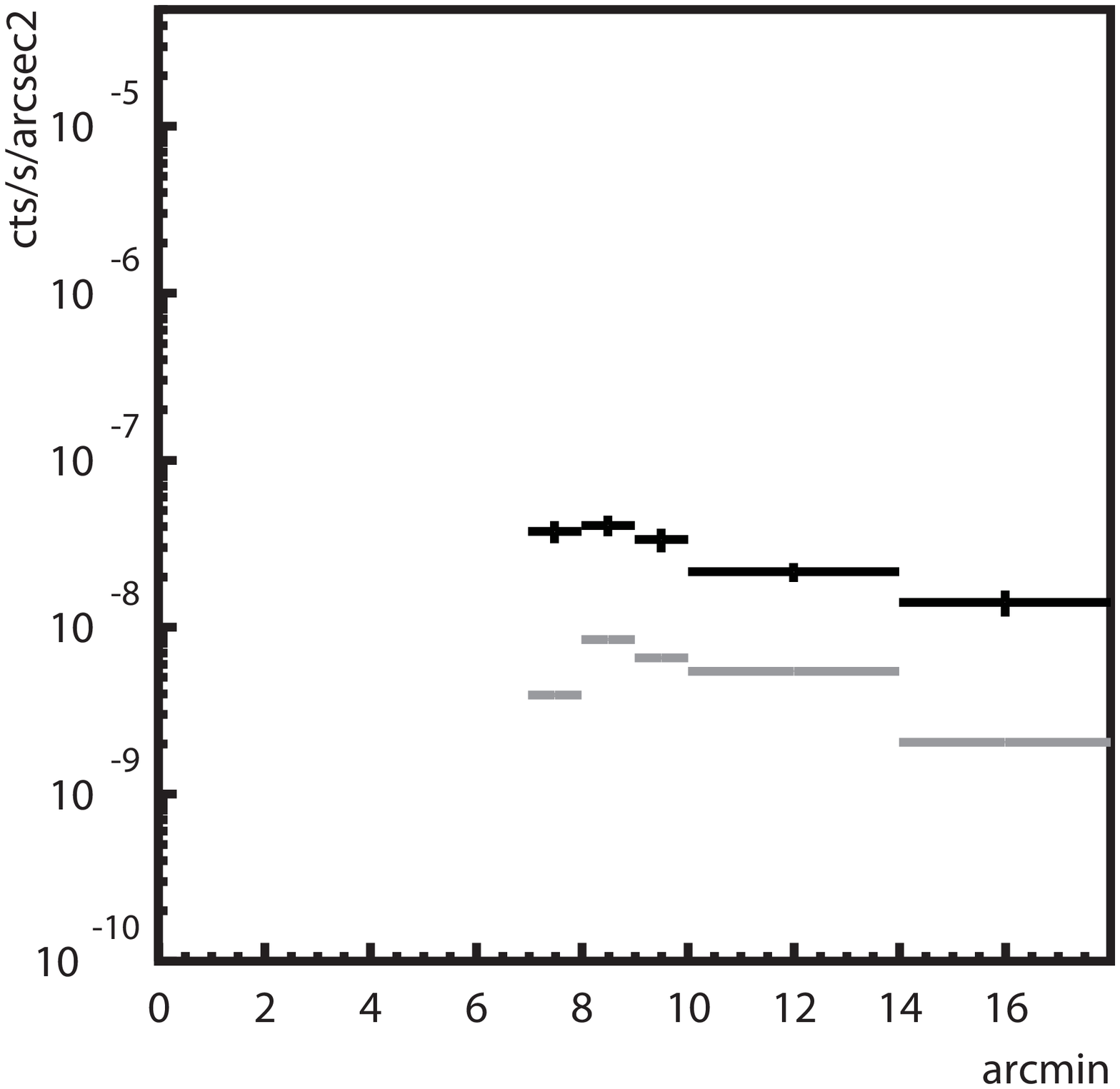} 
   \includegraphics[width=0.23 \textwidth,angle=0,clip]{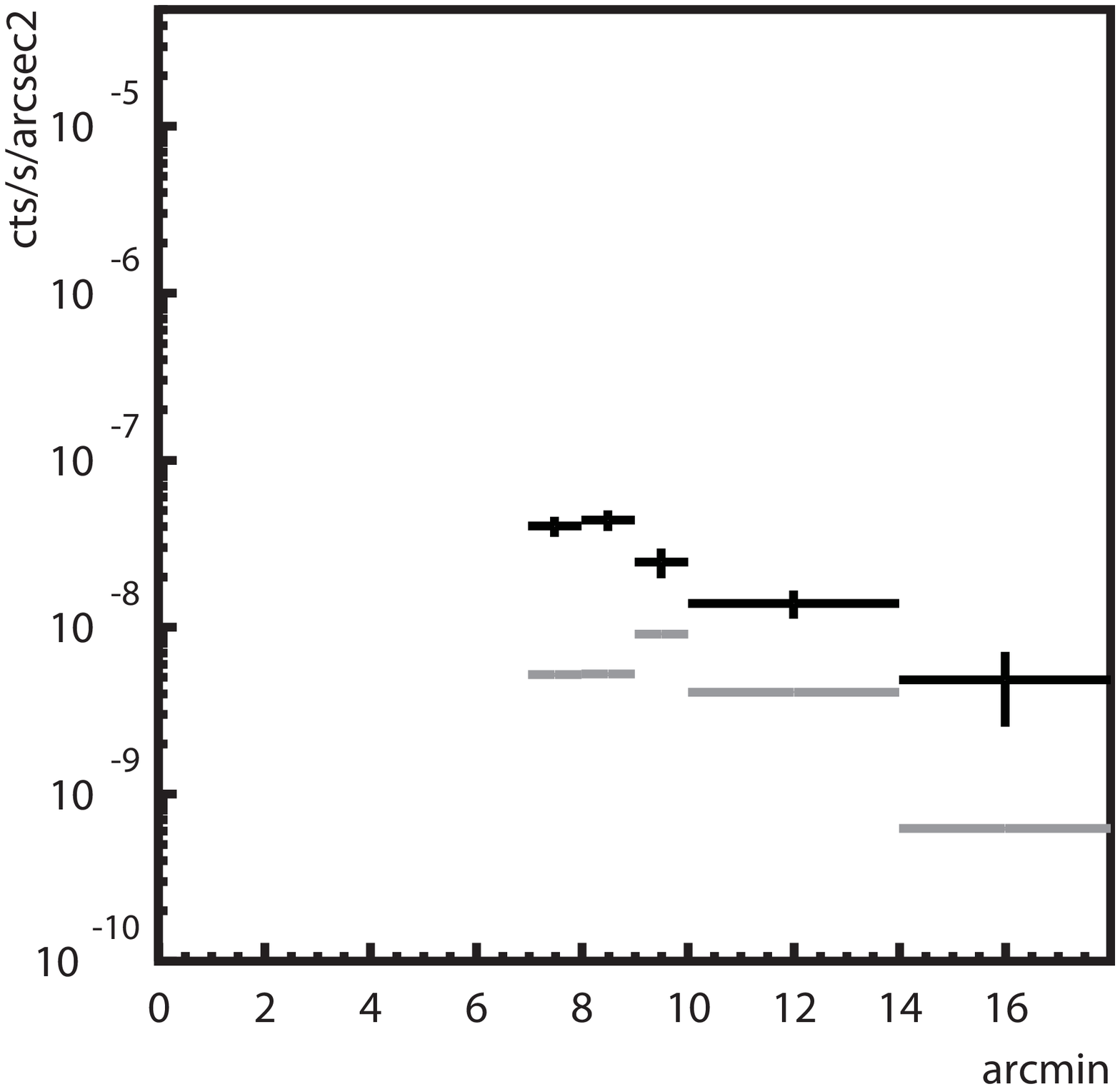} 
   \includegraphics[width=0.23 \textwidth,angle=0,clip]{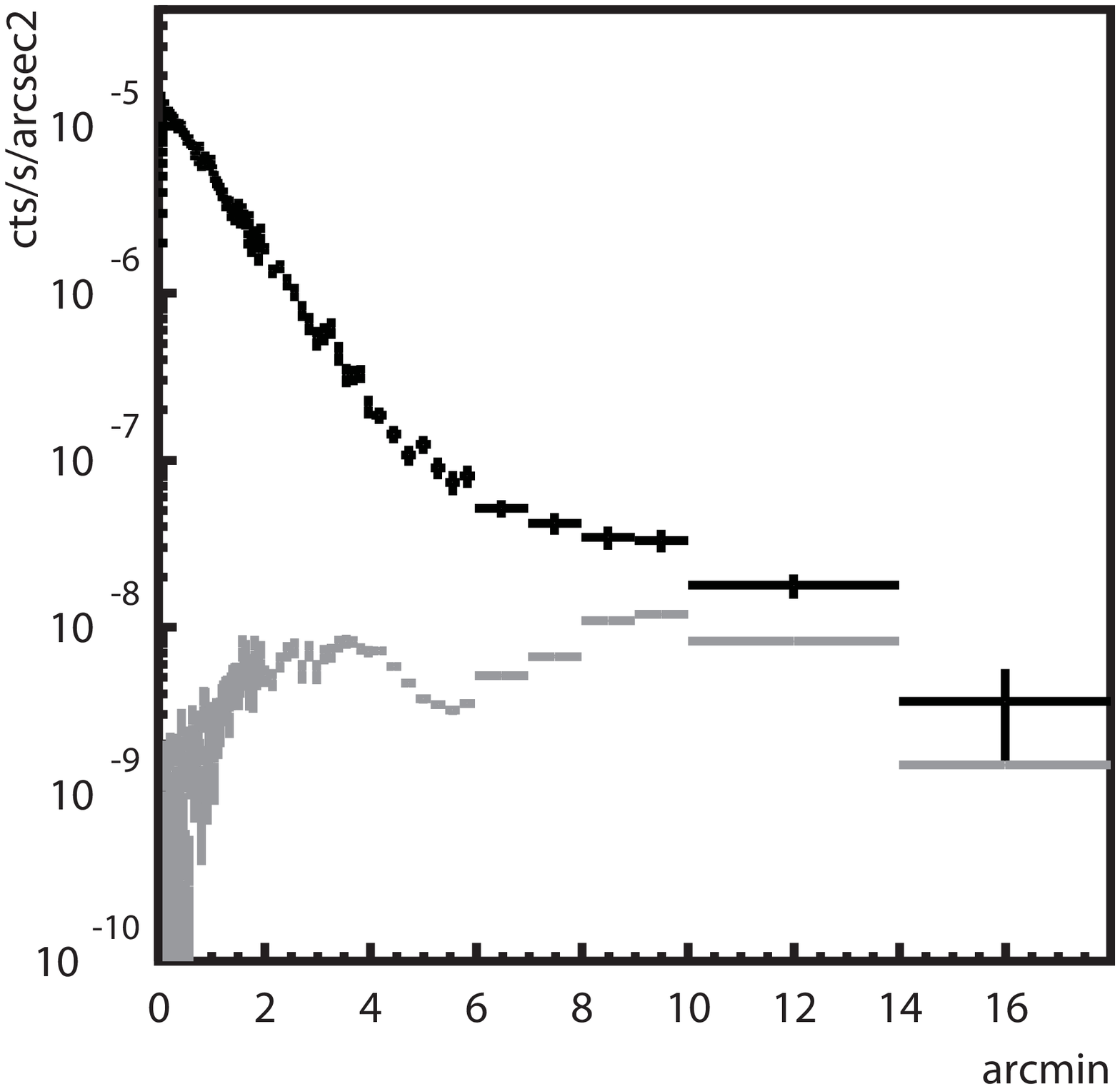} 
   \includegraphics[width=0.23 \textwidth,angle=0,clip]{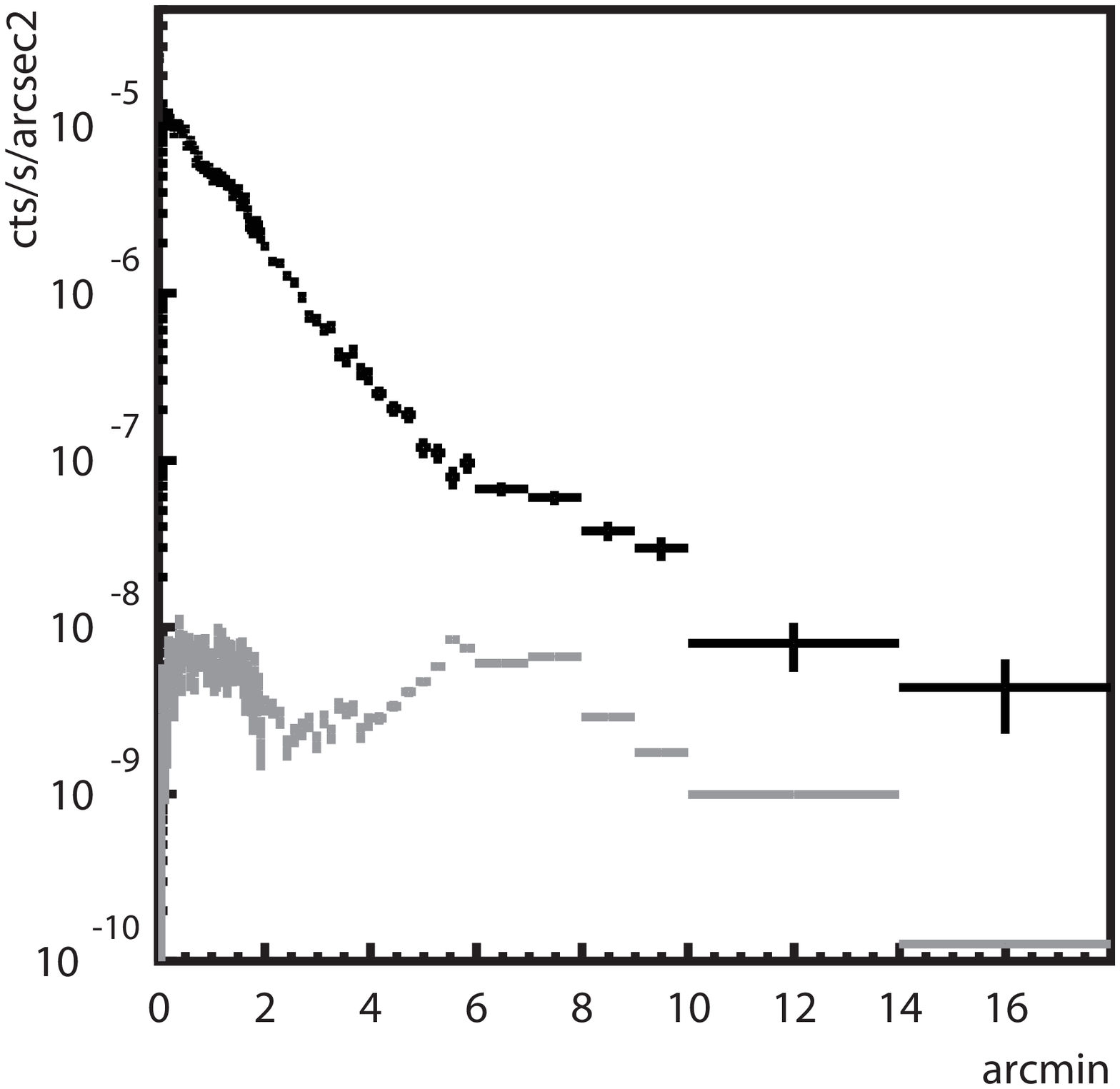} 
   \includegraphics[width=0.23 \textwidth,angle=0,clip]{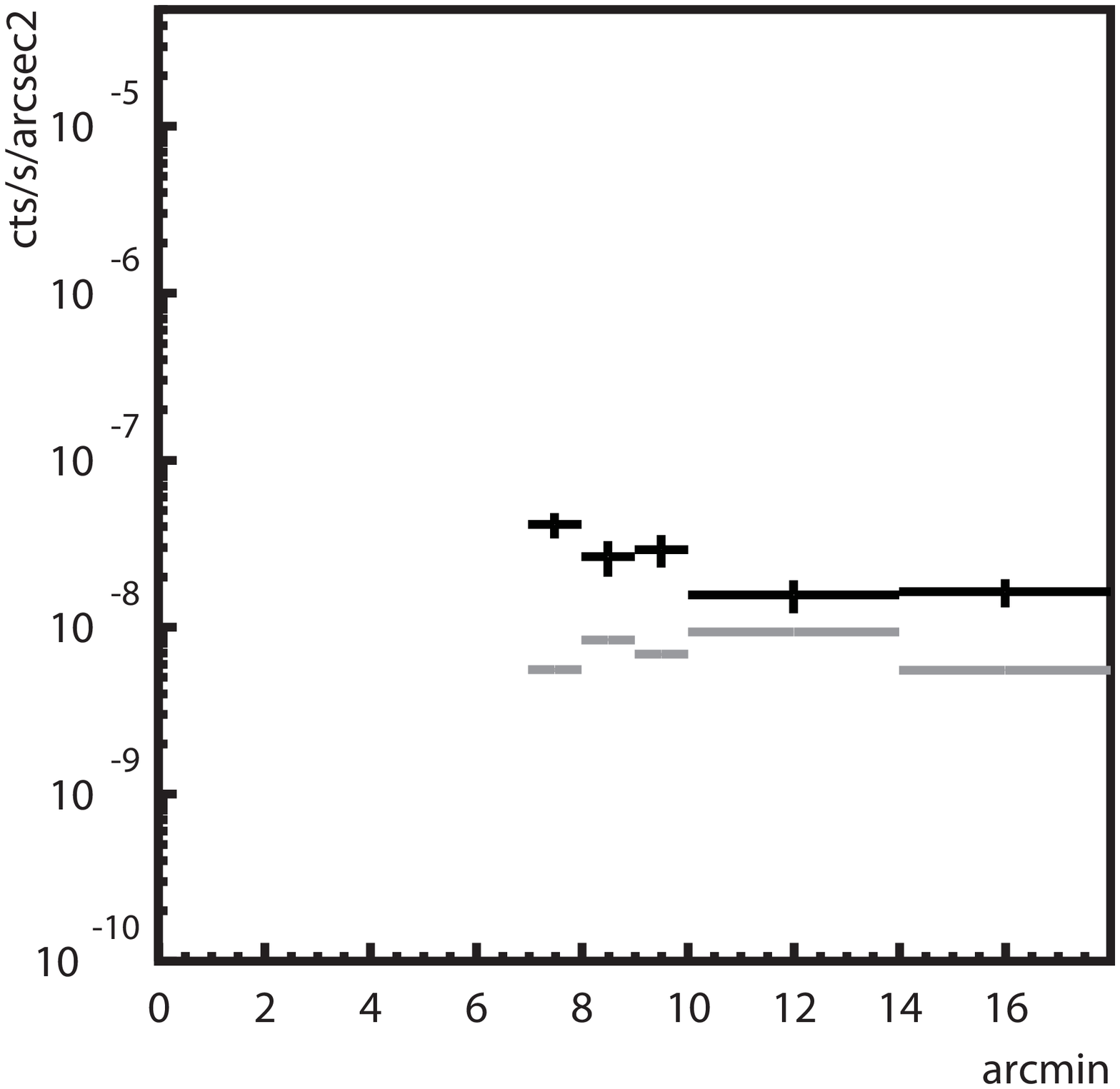} 
   \includegraphics[width=0.23 \textwidth,angle=0,clip]{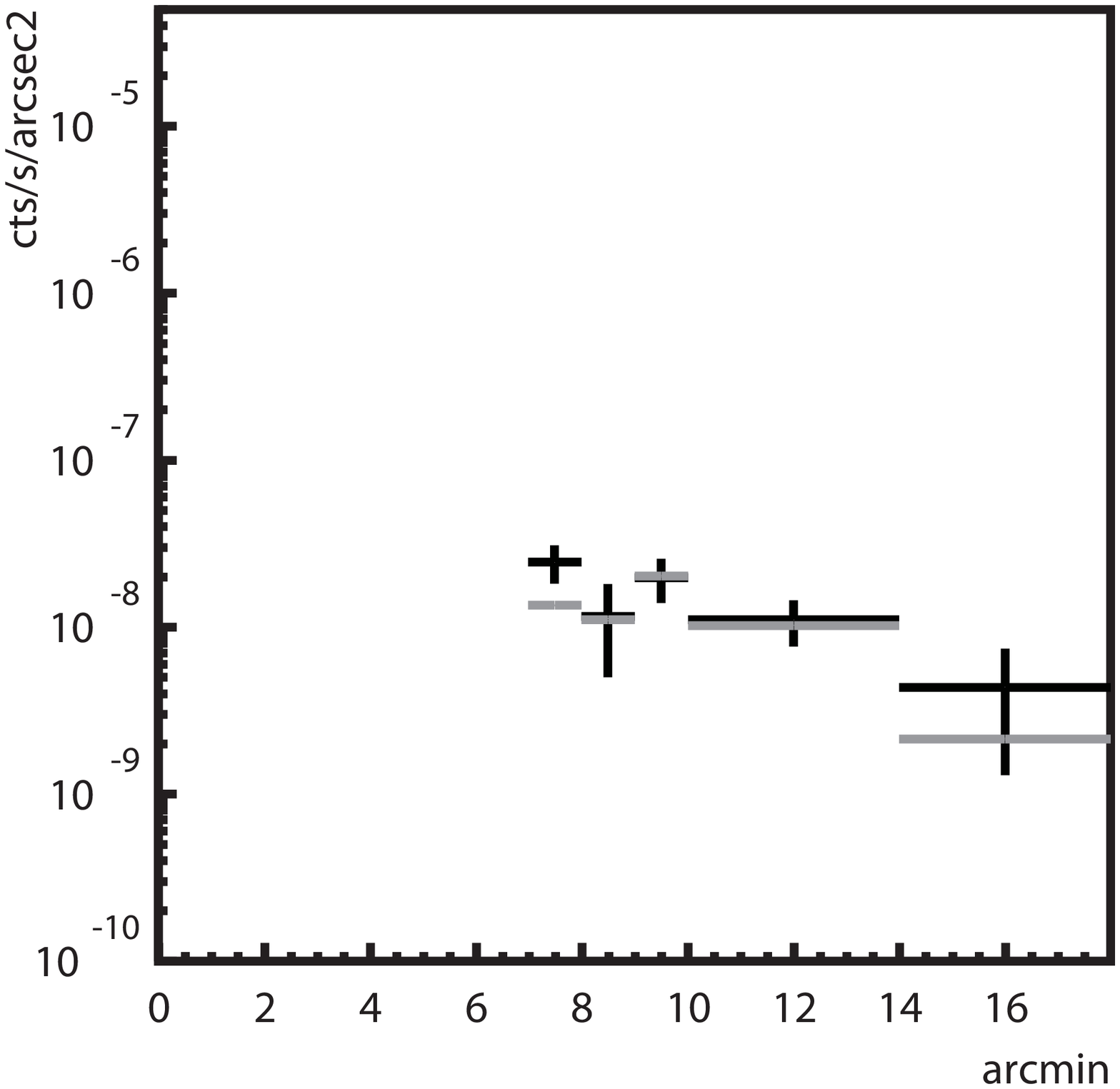} 
   \includegraphics[width=0.23 \textwidth,angle=0,clip]{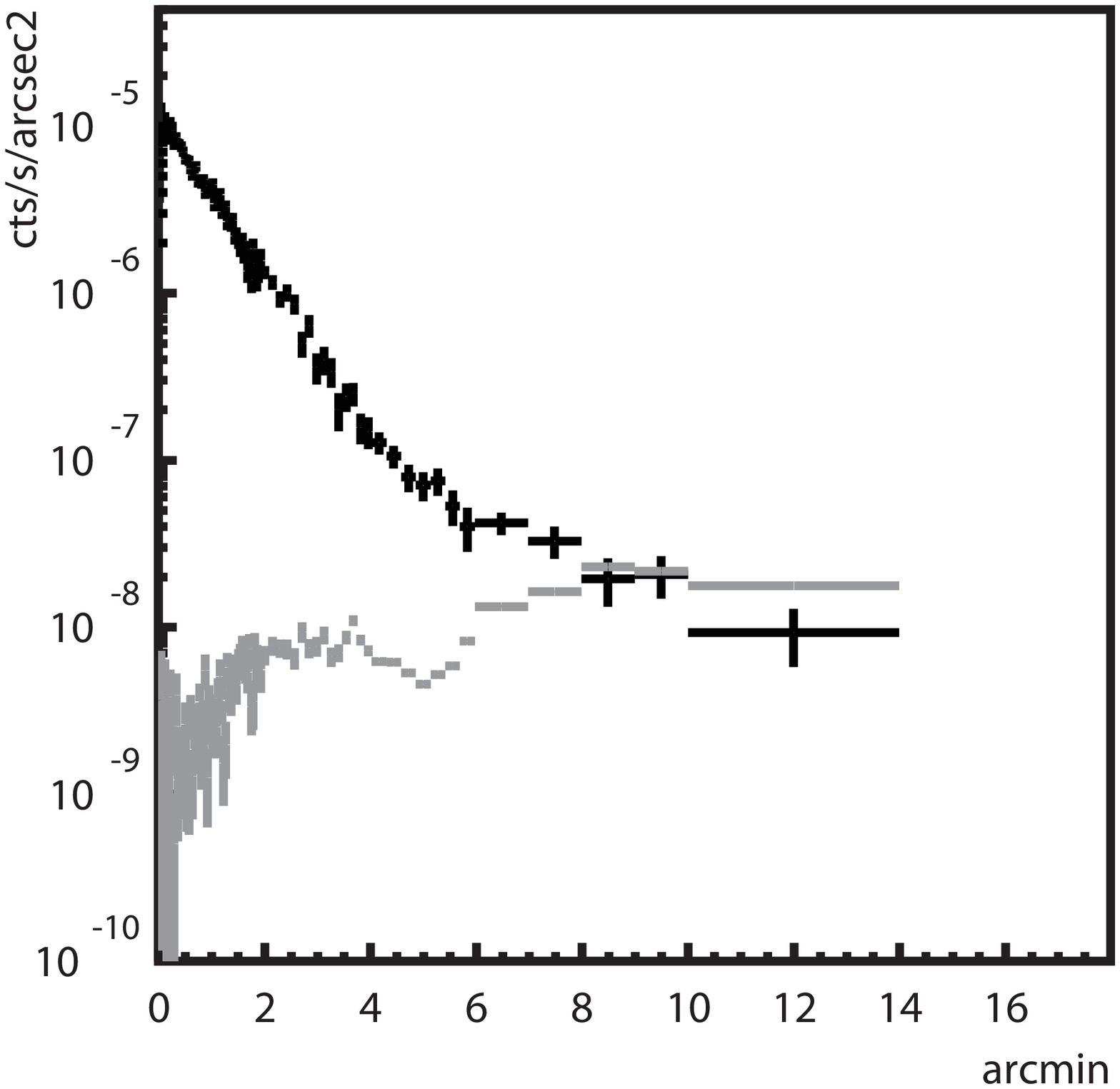} 
   \includegraphics[width=0.23 \textwidth,angle=0,clip]{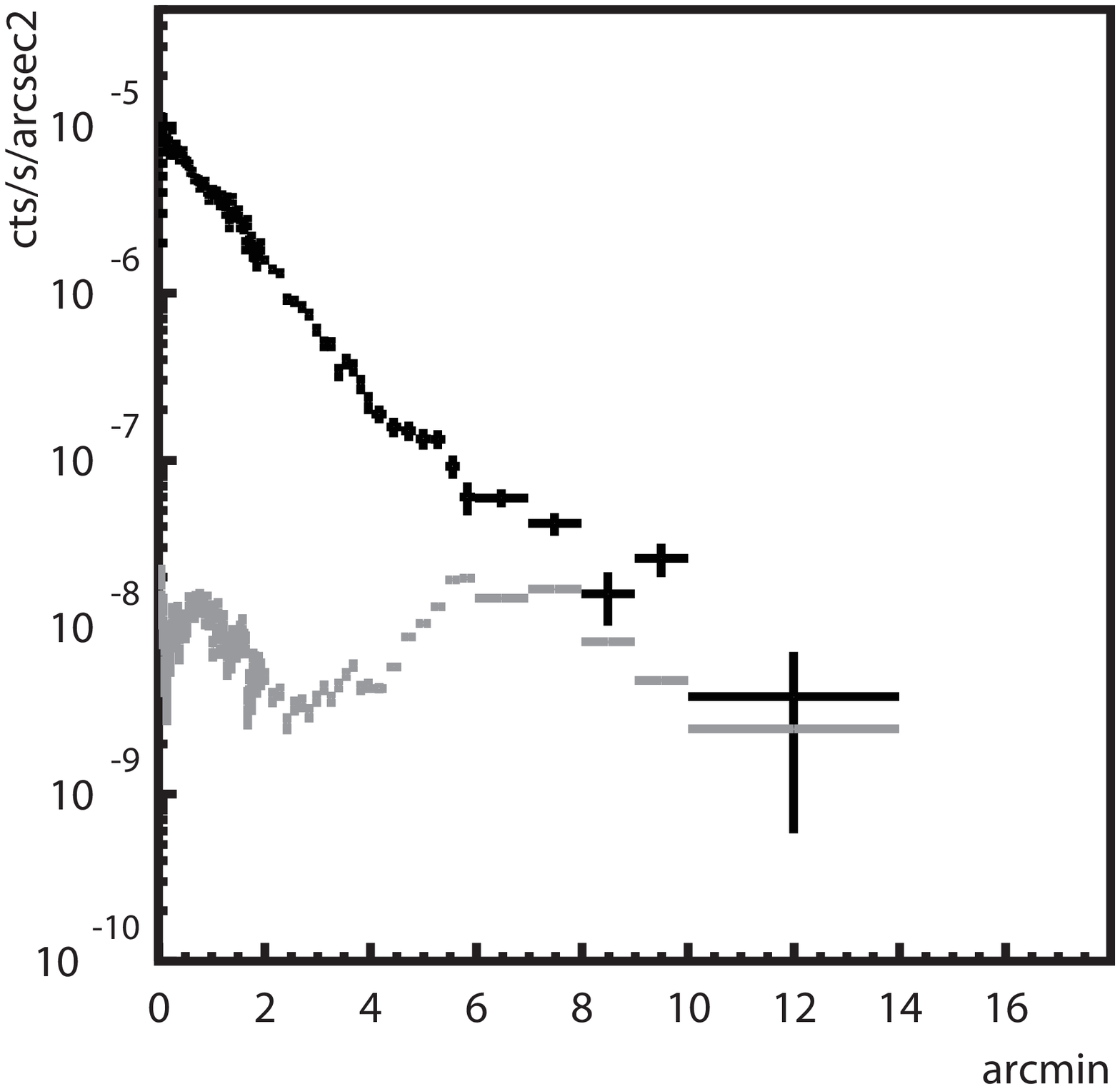} 
  \end{center}
  \caption{
(top) Simulated 0.5--2 keV profiles of residual point source signals (gray) in 
Offset1 through Offset4 pointings from left to right, 
shown with background-subtracted 0.5--2 keV  surface brightness profile of 
XIS-FI averaged over XIS0 and XIS3 (black).
In the error bars in this black profile, 1$\sigma$ uncertainties of the XRB and NXB
are added in quadrature to the corresponding statistical 1$\sigma$  errors. 
(bottom) the same as the top panels, but for 2--10 keV profiles. 
}
\label{fig:pntsrc_prof_azimuth}
\end{figure}

Using the derived crude 5-band spectra of the 49 point sources, we simulated XIS-FI (XIS0 + XIS3) 
images for the 4 observations using  \verb+xissim+. In order to improve statistics, we set
exposures which are 10 times longer than the actual observations. As shown in figure~\ref{fig:pntsrc_sim},
the whole area of XIS is contaminated with signals from point sources. 
Figure~\ref{fig:pntsrc_prof_azimuth} shows the simulated 0.5--2 keV and 2--10 keV profiles
of residual point-source signals in the four pointing directions.

\section{Parametric Density and Temperature Profiles}
\label{appen:n+Tpara}

\begin{figure}
  \begin{center}
   \includegraphics[width=0.6 \textwidth,angle=0,clip]{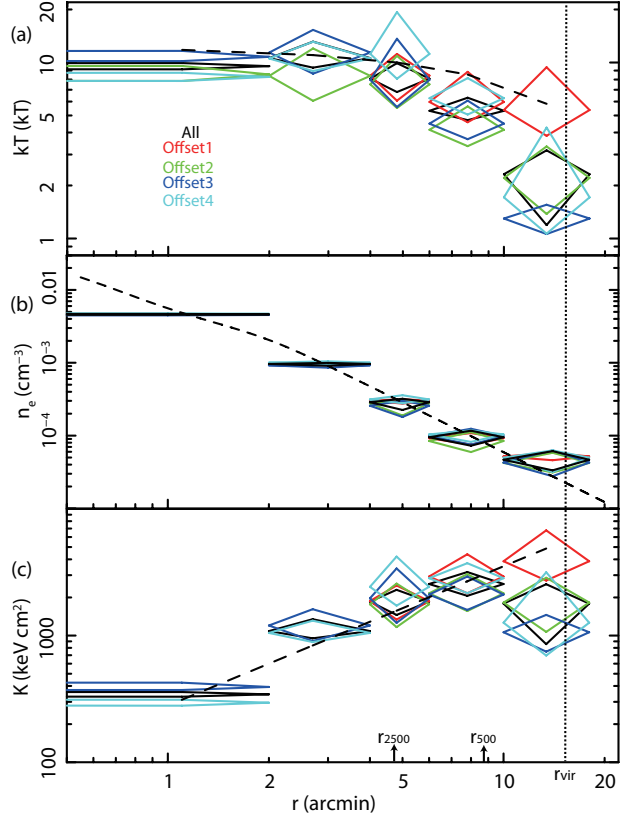} 
  \end{center}
  \caption{
(a) Projected temperature profiles obtained by spectral analyses of Offset1 (red), Offset2 (green),
Offset3 (blue), and Offset4 (cyan) observations. A temperature profile when 
all spectra of azimuthal regions at the same distance are summed
is also shown in black.
Dashed line is a 
temperature profile, $kT = 12.3 - 7.7\; (r/15\farcm6)$ keV, 
expected from the scaled temperature profile by \citet{pratt-2007}.
Dotted line shows the virial radius ($15\farcm6$). 
Arrows are the positions of $r_{\rm 500}$ ($8\farcm7$) and $r_{\rm 2500}$ ($4\farcm7$).
(b) the same as panel (a), but for deprojected electron number density profiles. 
See text for the detailed method of derivation (\S~\ref{subsection:ana:spectral}). 
Dashed line shows a model profile by \citet{peng-2009} (Model 3 in table 9). 
(c) the same as panel (a), but for entropy profiles
obtained by calculating $K = kT / n^{2/3}$ from profiles in panel (a) and (b).  
Dashed line shows $K \propto r^{1.1}$. Normalization of dashed line is 
arbitrarily scaled.
}
\label{fig:prof}
\end{figure}

\begin{figure*}
  \begin{center}
   \includegraphics[width=0.45 \textwidth,angle=0,clip]{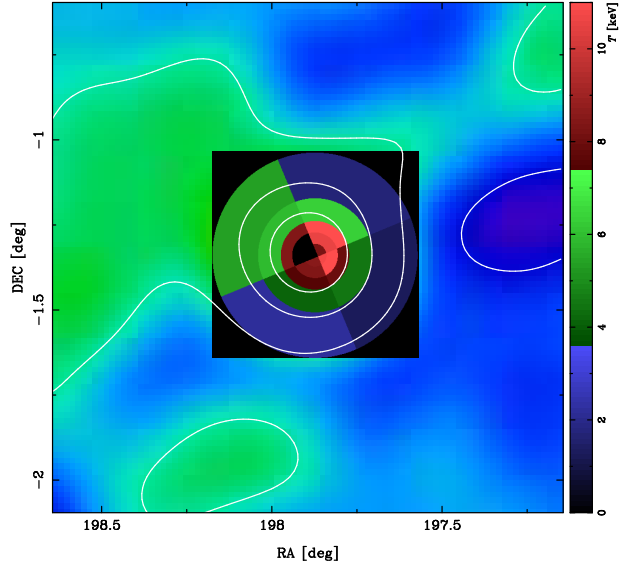} 
  \end{center}
  \caption{
  Gas temperature map of Abell~1689 (shown in the black box),
  embedded in the {\it SDSS} galaxy map which is a magnification of figure~\ref{fig:Mapsdss} (right). 
  Contours of the galaxy map are also overplotted for comparison. 
  The temperatures are the best-fit values determined by the spectral analysis in \S~\ref{subsection:ana:spectral}.
  See figure~\ref{fig:prof}(a) or table~\ref{table:fit-result} for their errors. 
}
\label{fig:Map_kt_sdss}
\end{figure*}

We parametrized the electron density and temperature profiles shown in figure \ref{fig:prof}, which is necessary to measure the hydrostatic equilibrium mass (\S \ref{subsec:ICM_outskirts}) as well as to conduct a joint analysis of {\it Suzaku} X-ray satellite and {\it Subaru/Suprime-Cam}+{\it HST/ACS} lensing analysis (\S \ref{sec:joint}). 

The $\beta$-model \cite[e.g.][]{cavaliere-1976} or its generalized version \cite[]{vikhlinin-2009}
are often used to express electron density profiles in clusters.
Since the low angular resolution of the {\it Suzaku} satellite does not
require the complicated modeling \cite[]{vikhlinin-2009}, the analytic
expression we use for deprojected electron density profile is given by  
\begin{eqnarray}
n_e(r)=n_{e,0}\left(1+\left(\frac{r}{r_{c,1}}\right)^2\right)^{-3\beta/2}\left(1+\left(\frac{r}{r_{c,2}}\right)^2\right),\label{eq:ne_pr}
\end{eqnarray}
where $n_{e,0}$ is the central electron number density, $\beta$ is the
slope parameter, and $r_{c,1}$ and $r_{c,2}$ are the core radii. 
In this modified expression of $\beta$ model, 
the number density of the outskirts $r>10\farcm0$ are much better fitted with
the outer radius $r_{c,2}$, than a single $\beta$ model.
Unfortunately, since the number of the data points is small, the core
radius, $r_{c,2}$, is not constrained well. We therefore fixed
$r_{c,2}=3r_{c,1}$. 
Our results (\S \ref{section:dis}) are not significantly changed by choosing a constant
factor of $r_{c,2}/r_{c,1}$. 
The best-fit parameters for all, Offset1 and Offset234 regions are summarized in table \ref{table:nfit}.

The observed projected temperature profile is described using a simple parametrization
\begin{eqnarray}
 T(r)=T_0
  \left(\frac{r}{r_t}\right)^{1/2}\left[1+\left(\frac{r}{r_t}\right)^2\right]^{-\alpha}, \label{eq:T_pr}
\end{eqnarray}
with the temperature normalization $T_0$, the scale radius $r_t$ and the slope parameter $\alpha$.
We note that the assumed inner slope $1/2$ is insignificant for our main purpose to describe the temperature profile in the outskirts.

We fit the density and temperature profiles in the whole region of the cluster, 
Offset1 and Offset234 regions with the above two models, respectively. 
In the fitting of Offset1 region, we utilize the data of the other
region instead of lack of the first two bins in the profiles. 
The best-fit parameters for all, Offset1 and Offset234 regions are summarized in table \ref{table:nfit}.

\newpage


\begin{figure}
  \begin{center}
   \includegraphics[width=0.4\textwidth,angle=0,clip]{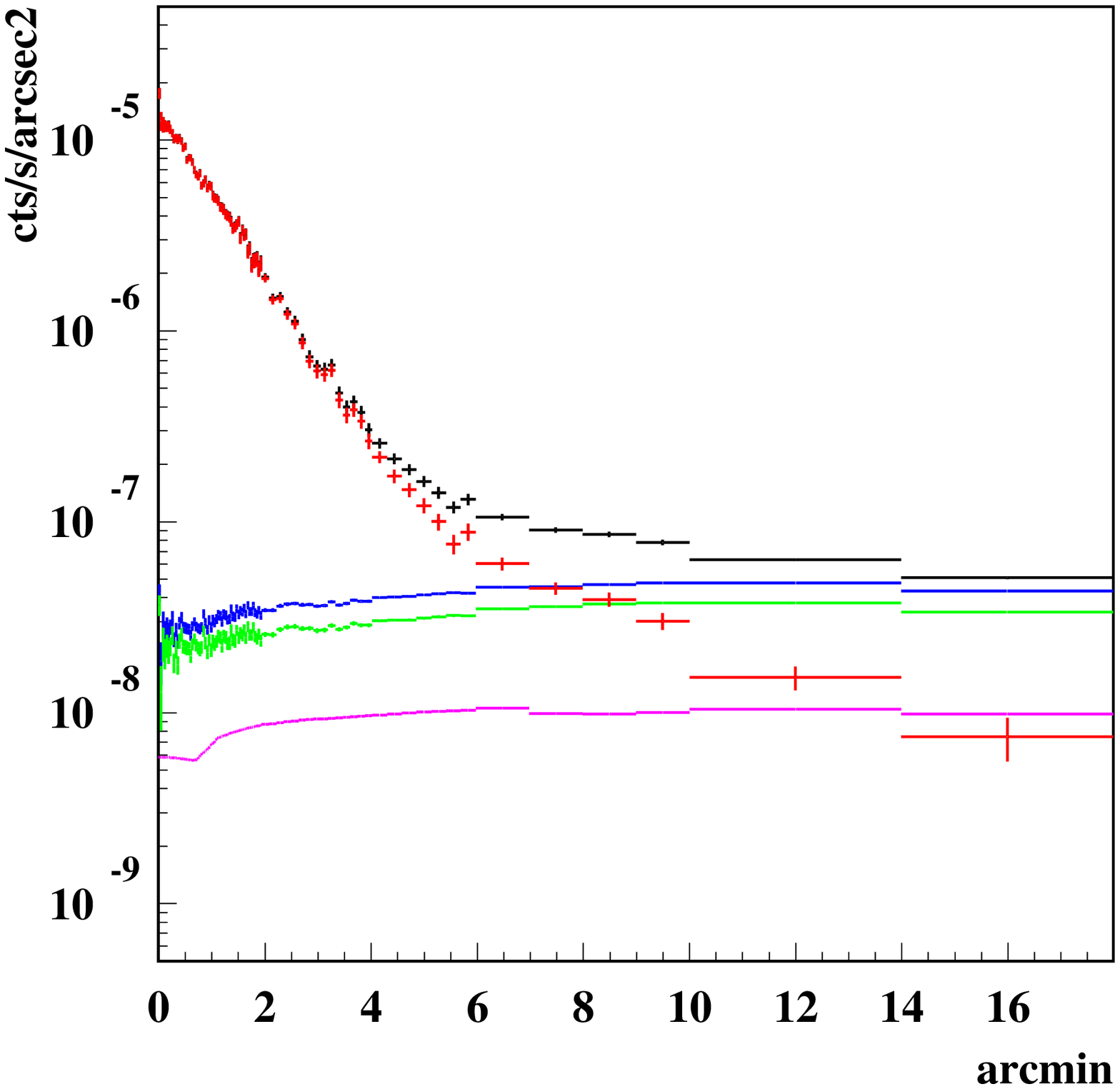} 
   \includegraphics[width=0.4\textwidth,angle=0,clip]{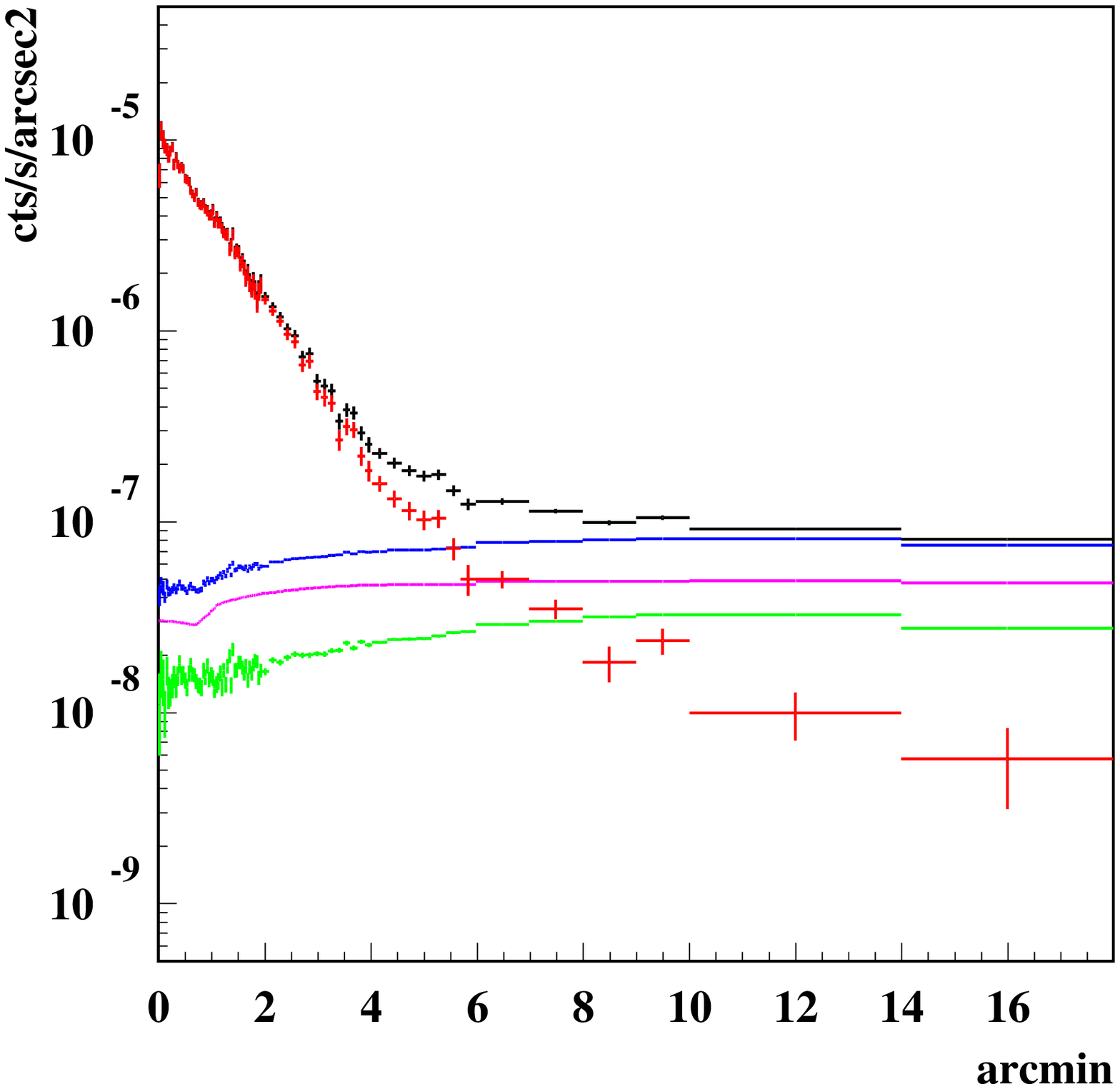} 
  \end{center}
  \caption{(left) Surface brightness profiles made of 0.5--2 keV XIS-FI (averaged over XIS0 and XIS3)
image, using all the Offset1 through Offset4 observations. Point sources specified in 
figure~\ref{fig:mosaic} (right) are removed, but vignetting is not corrected.
Raw (background inclusive) profile (black) is shown with XRB (green), 
NXB (magenta), and XRB$+$NXB (blue) profiles. 
A resulting background-subtracted profile (black $-$ blue) is shown in red. 
The error bars in these profiles are 1$\sigma$. 
In the error bars in the red profile, 1$\sigma$ uncertainties of the XRB and NXB
are added in quadrature to the corresponding statistical 1$\sigma$  errors. 
(right) the same as the left panel, but for 2--10 keV profiles. 
}
\label{fig:prof_all}
\end{figure}

\clearpage

\begin{deluxetable}{lccc}
\tablecaption{ {\it Suzaku} Observation log of Abell 1689. \label{table:observation} }
\tablewidth{0pt}
\tablehead{
\colhead{Observation (ID)} & \colhead{Start\tablenotemark{a}} & 
\colhead{End\tablenotemark{a}} & \colhead{}Exposure\tablenotemark{b} 
}
\startdata
Offset1 (803024010)& 2008/July/23 04:42:40 & 2008/July/24 08:22:23 & 39.2 \\
Offset2 (803025010)& 2008/July/24 08:23:04 & 2008/July/25 10:09:18 & 38.2 \\
Offset3 (803026010)& 2008/July/25 10:09:59 & 2008/July/26 11:23:19 & 39.3 \\
Offset4 (803027010)& 2008/July/26 11:24:00 & 2008/July/27 12:13:14 & 37.9 \\
\enddata
\tablenotetext{a}{Time is shown in UT.}
\tablenotetext{b}{Effective XIS exposure in units of ks, obtained after the event screenings.}
\end{deluxetable}
\begin{table}
 \caption{RASS averaged count rates around Abell~1689 and the blank-sky fields.}
\label{table:rass-cnt}
   \begin{center}
    \begin{tabular}{lccc}
\tableline 
\tableline 
Band\tablenotemark{a} & $0.5^{\circ}$--$1.0^{\circ}$\tablenotemark{b} & Q1334\tablenotemark{c} & N4636\_GAL\tablenotemark{c} \\
\tableline 
1/4 keV  & $858.3 \pm 36.0$   & $816.1 \pm 28.0$  & $1072.1 \pm 42.1$\\
3/4 keV  & $157.6 \pm 15.8$   & $133.3 \pm 12.0$   & $238.3 \pm 20.1$\\
1.5 keV  & $139.1 \pm 14.3$   & $132.2 \pm 12.0$   & $159.2 \pm 17.8$\\
\tableline 
\end{tabular}
\end{center}
\tablenotetext{a}{RASS 1/4 keV, 3/4 keV, and 1.5 keV bands correspond to
0.12--0.284 keV, 0.47--1.21 keV, and 0.76--2.04 keV, respectively.} 
\tablenotetext{b}{RASS count rate in an annulus around Abell~1689, in units of 10$^{-6}$ counts s$^{-1}$ arcmin$^{-2}$.}
\tablenotetext{c}{RASS count rate averaged over the HLD field 
($1^{\circ}$ radius) in units of 10$^{-6}$ counts s$^{-1}$ arcmin$^{-2}$.} 
\end{table}
\begin{table}
 \caption{Results of model fittings to XIS spectra of the blank sky fields.}
\label{table:fit-bgd}
   \begin{center}
    \begin{tabular}{cccc}
\tableline 
\tableline 
$\Gamma$(CXB) & $S_{\rm x}$(CXB)\tablenotemark{a} & $S_{\rm x}$(GFE)\tablenotemark{b} & $\chi^2/$dof \\
\tableline 
1.412 (fix)& $6.3^{+0.2}_{-0.2} \times 10^{-8}$& $3.5^{+1.2}_{-1.2} \times 10^{-8}$ &  162.4/127 \\
\tableline 
\end{tabular}
\end{center}
\tablenotetext{a}{The 2.0--10.0 keV CXB surface brightness in units of erg s$^{-1}$ cm$^{-2}$ sr$^{-1}$.}
\tablenotetext{b}{The 0.3--1.0 keV GFE surface brightness in units of erg s$^{-1}$ cm$^{-2}$ sr$^{-1}$.} 
\end{table}

\begin{table}
 \caption{Significance of the signal outside $10'$.}
\label{table:significance}
   \begin{center}
    \begin{tabular}{ccc}
\tableline 
\tableline 
 & Fraction\tablenotemark{a} & Significance\tablenotemark{b} \\
\tableline 
Offset1 0.5--2 keV & 0.21 & 5.4$\sigma$ \\
Offset2 0.5--2 keV & 0.25 & 2.9$\sigma$ \\
Offset3 0.5--2 keV & 0.45 & 2.0$\sigma$ \\
Offset4 0.5--2 keV & 0.09 & 2.6$\sigma$ \\
\tableline 
Offset1 2--10 keV & 0.47 & 2.1$\sigma$ \\
Offset2 2--10 keV & 0.80 & 0.0$\sigma$ \\
Offset3 2--10 keV & $> 1$ & 0.0$\sigma$ \\
Offset4 2--10 keV & 0.64 & 0.2$\sigma$ \\
\tableline 
\end{tabular}
\end{center}
\tablenotetext{a}{Fraction of the residual point source signal defined as 
(simulated signal of point sources)/(background-subtracted signal)
outside $10'$.}
\tablenotetext{b}{Significance of signal outside $10'$, after subtracting 
contributions from point sources and inter ICM emission. As to the latter contribution, constant
value of 22\% is used for all the pointings and energy ranges.} 

\end{table}


\begin{deluxetable}{cccccccccc}
\tabletypesize{\scriptsize}
\tablecaption{Fitting results of spectral analysis for the five annular regions. \label{table:fit-result}}
\tablewidth{0pt}
\tablehead{
\colhead{ Region} & \colhead{ $kT_1$\tablenotemark{a}} & \colhead{ $Z_1$\tablenotemark{a}} & \colhead{ $kT_2$\tablenotemark{a}} & \colhead{ $Z_2$\tablenotemark{a}} &
\colhead{ $kT_3$\tablenotemark{a}} & \colhead{ $Z_3$\tablenotemark{a}} & \colhead{ $kT_4$\tablenotemark{a}} &
\colhead{ $Z_4$\tablenotemark{a}} & \colhead{ $\chi^2$/D.O.F\tablenotemark{b}}  
}
\startdata
 $0'$--$2'$ &  --  &  -- &  $8.55^{+1.01}_{-0.70}$ &  $0.36^{+0.15}_{-0.15}$ &  $10.76^{+0.88}_{-0.58}$ &  $0.34^{+0.09}_{-0.09}$ &  $8.23^{+0.46}_{-0.42}$ &  $0.26^{+0.06}_{-0.06}$ &  $752.0/605$  \\  
  $2'$--$4'$ &  --  & -- & $8.39^{+3.63}_{-2.33}$ & $0.12^{+0.98}_{-0.12}$ & $11.39^{+3.92}_{-2.75}$ & $0.77^{+0.00}_{-0.77}$ & $10.53^{+2.60}_{-1.59}$ & $0.49^{+0.26}_{-0.25}$  & $273.2/204$ \\  
  $4'$--$6'$ &  $8.42^{+2.76}_{-2.35}$ & $0.77^{+0.00}_{-0.77}$ & $7.51^{+3.40}_{-2.00}$ & $3.76^{+0.00}_{-3.76}$ & $7.96^{+5.68}_{-2.38}$ & $0.92^{+0.00}_{-0.92}$ & $11.19^{+8.15}_{-3.11}$ & $0.94^{+1.19}_{-0.94}$ & $319.0/210$ \\ 
  $6'$--$10'$ &  $5.96^{+2.89}_{-1.39}$ & $0.71^{+1.11}_{-0.71}$ & $4.14^{+1.48}_{-0.80}$ &  $1.40^{+2.30}_{-1.17}$ & $4.49^{+1.58}_{-0.83}$ & $1.21^{+1.65}_{-1.02}$ & $6.27^{+1.88}_{-1.21}$ & $0.89^{+0.99}_{-0.72}$ & $394.3/262$ \\
  $10'$--$18'$ &  $5.37^{+4.02}_{-1.54}$ & $0.64^{+1.07}_{-0.64}$ & $2.21^{+1.13}_{-0.83}$ & $0.49^{+1.94}_{-0.49}$ & $1.30^{+0.26}_{-0.24}$ & $0.50^{+1.85}_{-0.38}$ & $1.71^{+2.56}_{-0.65}$ & $3.65^{+0.00}_{-3.65}$ & $665.7/458$ \\
\tableline
Region & \multicolumn{4}{c}{$kT_{\rm all}$\tablenotemark{a}}  & \multicolumn{4}{c}{$Z_{\rm all}$\tablenotemark{a}} & $\chi^2$/D.O.F\tablenotemark{c}  \\
\tableline
 $0'$--$2'$ &  \multicolumn{4}{c}{$9.53^{+0.39}_{-0.39}$}    & \multicolumn{4}{c}{$0.31^{+0.05}_{-0.05}$} &  $822.5/611$\\
 $2'$--$4'$ &    \multicolumn{4}{c}{$10.58^{+2.55}_{-1.24}$} & \multicolumn{4}{c}{$0.63^{+0.30}_{-0.31}$} & $499.2/210$ \\
 $4'$--$6'$ &    \multicolumn{4}{c}{$8.09^{+1.82}_{-1.30}$ } & \multicolumn{4}{c}{$0.65^{+0.66}_{-0.60}$} & $429.7/219$\\
 $6'$--$10'$ &   \multicolumn{4}{c}{$5.31^{+1.00}_{-0.61}$ } & \multicolumn{4}{c}{$0.90^{+0.58}_{-0.48}$} & $423.0/271$  \\
 $10'$--$18'$ &   \multicolumn{4}{c}{$2.32^{+0.85}_{-1.13}$ }& \multicolumn{4}{c}{$0.35^{+0.65}_{-0.31}$} & $743.6/467$
\enddata
\tablenotetext{a}{$kT_1$, $kT_2$, $kT_3$, $kT_4$, and $kT_{\rm all}$ represent for temperatures of Offset1, Offset2, Offset3, Offset4 and the sum of all the azimuthal regions, respectively, in units of keV. Similarly, $Z_1$, $Z_2$, $Z_3$, $Z_4$, and $Z_{\rm all}$ represent for metal abundances, in units of solar abundance of \citet{anders-1989}.}
\tablenotetext{b}{Reduced $\chi^2$ of the fit when parameters are separately determined for the four offset pointings.}
\tablenotetext{c}{Reduced $\chi^2$ of the fit when parameters are tied between the four offset pointings.}
\end{deluxetable}

\begin{deluxetable}{cccccc}
\tabletypesize{\scriptsize}
\tablecaption{Deprojected electron number densities of the five annular regions. \label{table:density}}
\tablewidth{0pt}
\tablehead{
\colhead{Region} & \colhead{$n_{\rm e,1}$\tablenotemark{a}} & \colhead{$n_{\rm e,2}$\tablenotemark{a}} &
\colhead{$n_{\rm e,3}$\tablenotemark{a}} & \colhead{$n_{\rm e,4}$\tablenotemark{a}} & \colhead{$n_{\rm e,all}$\tablenotemark{a}} 
}
\startdata
$0'$--$2'$   & --                                  &--                                   & $ 4.53^{+0.06}_{-0.06}\times 10^{-3}$ &  $4.70^{+0.07}_{-0.07}\times 10^{-3}$ & $4.62^{+0.07}_{-0.07}\times 10^{-3}$ \\ 
$2'$--$4'$   & --                                  &--                                   & $ 9.23^{+0.24}_{-0.67}\times 10^{-4}$ &  $1.00^{+0.04}_{-0.04}\times 10^{-3}$ & $9.64^{+0.32}_{-0.56}\times 10^{-4}$ \\
$4'$--$6'$   & $3.02^{+0.14}_{-0.27}\times 10^{-4}$  & $2.77^{+0.21}_{-0.88}\times 10^{-4}$  & $ 2.57^{+0.44}_{-0.77}\times 10^{-4}$ &  $3.14^{+0.44}_{-0.36}\times 10^{-4} $ & $2.88^{+0.33}_{-0.63}\times 10^{-4}$ \\
$6'$--$10'$  & $9.21^{+1.63}_{-1.66}\times 10^{-5}$  & $8.44^{+2.74}_{-2.48}\times 10^{-5}$  & $ 9.86^{+2.55}_{-2.33}\times 10^{-5}$ &  $1.04^{+0.14}_{-0.22}\times 10^{-4} $ & $9.49^{+2.20}_{-2.22}\times 10^{-5}$ \\
$10'$--$18'$ & $5.18^{+0.62}_{-0.59}\times 10^{-5}$  & $4.22^{+1.61}_{-1.10}\times 10^{-5}$  & $ 4.27^{+2.01}_{-1.50}\times 10^{-5}$ &  $4.99^{+1.28}_{-1.82}\times 10^{-5} $ & $4.66^{+1.47}_{-1.33}\times 10^{-5}$ \\
\enddata
\tablenotetext{a}{$n_{\rm e, 1}$, $n_{\rm e, 2}$, $n_{\rm e, 3}$,  $n_{\rm e, 4}$, and $n_{\rm e, all}$ represent for electron number densities of Offset1, Offset2, Offset3, Offset4, and the sum of all the azimuthal regions, respectively, in units of cm$^{-3}$.}
\end{deluxetable}


\begin{deluxetable}{ccccccc}
\tabletypesize{\scriptsize}
\tablecaption{Best-fit parameters for electron density and temperature profiles. \label{table:nfit}}
\tablewidth{0pt}
\tablehead{
\colhead{Region} 
& \colhead{$n_{\rm e,0}$\tablenotemark{a}} 
& \colhead{$\beta$\tablenotemark{b}}
& \colhead{$r_{\rm c,1}$\tablenotemark{c}}
& \colhead{$T_{\rm 0}$\tablenotemark{d}}
& \colhead{$\alpha$\tablenotemark{e}} 
& \colhead{$r_{\rm t}$\tablenotemark{f}} 
}
\startdata
All   & $(0.98\pm0.04)\times10^{-2}$
      & $1.16\pm0.02$
      & $1.43\pm0.06$
      & $19.39\pm0.61$
      & $1.03\pm0.07$ 
      & $3.88\pm0.29$\\
Offset1   & $(1.01\pm0.11)\times10^{-2}$
      & $1.13\pm0.03$
      & $1.36\pm0.13$
      & $15.85\pm1.30$
      & $0.60\pm0.10$ 
      & $2.42\pm0.53$ \\
Offset234   & $(0.97\pm0.04)\times10^{-2}$
      & $1.17\pm0.02$
      & $1.44\pm0.06$
      & $21.10\pm1.32$
      & $1.32\pm0.19$ 
      & $4.72\pm0.69$ \\
\enddata
\tablenotetext{a-c}{$n_{\rm e, 0}$, $\beta$ and $r_{\rm c,1}$ are 
the central electron density (cm$^{-3}$), the slope, the core
 radius (arcmin) for the parametrized density profile
 (eq. \ref{eq:ne_pr}), respectively}
\tablenotetext{d-f}{$T_0$, $\alpha$, $r_{\rm t}$ are the normalized
 temperature (keV), the slope, the scale radius (arcmin)
for the  parametrized temperature profile (eq. \ref{eq:T_pr}), respectively.}
\end{deluxetable}

\begin{table}
 \caption{Lensing Mass Parameters}
\label{table:lensmass}
   \begin{center}
    \begin{tabular}{ccc}
\hline
\hline
Model  & $M_{\rm vir}$\tablenotemark{a}
        & $C_{\rm vir}$\tablenotemark{b} \\
        & $(10^{15}h^{-1}M_\sun)$
        & \\ 
\hline
NFW\tablenotemark{c}     &  $1.47^{+0.59}_{-0.33}$
        &  $12.7\pm2.9$ \\
Spherical\tablenotemark{d} & $1.30_{-0.39}^{+0.38}$
        & -\\
NFW-equivalent triaxial model\tablenotemark{e} & $1.26^{+0.28}_{-0.49}$
      & $16.9^{+2.2}_{-12.9}$\\
\hline
\end{tabular}
\end{center}
\tablenotetext{a}{Virial mass.}
\tablenotetext{b}{Concentration parameter of the NFW profile.}
\tablenotetext{c}{NFW model parameters taken from
 \cite{umetsu-2008}. The quoted errors include both statistical and
 systematic uncertainties.}
\tablenotetext{d}{Deprojected spherical mass.}
\tablenotetext{e}{NFW-equivalent triaxial model parameters taken from
 \cite{oguri-2005}.}

\end{table}


\begin{table}
\caption{Point sources identified in XIS FOVs of Offset1 through Offset4 observations. \label{table:pntsrc}}
\begin{center}
 \begin{tabular}{lccc}
\hline
\hline
\#\tablenotemark{a}  &
$\alpha$\tablenotemark{b}  &
$\delta$\tablenotemark{b}  &
Flux\tablenotemark{c} \\
\hline
 %
 1$^{*}$   &198.03427  &-1.1213164   &3.09405e-14  \\
 2$^{*}$    &198.05514  &-1.5692276   &2.44946e-14  \\
 3$^{*}$    &198.02596  &-1.5233972   &2.57838e-14  \\
 4   &197.82543  &-1.2416705   &1.10009e-13  \\
 5   &197.78077  &-1.4828847   &7.7066e-14   \\
 6   &197.79435  &-1.5023029   &8.81786e-14  \\
 7   &197.98643  &-1.2203908   &4.71844e-14  \\
 8   &197.76312  &-1.4169539   &9.32276e-14  \\
 9   &197.81012  &-1.5273204   &1.33557e-13  \\
 10  &197.87293  &-1.2675864   &2.20443e-14  \\
 11  &197.98223  &-1.4325242   &3.78360e-14  \\
 12  &197.93268  &-1.5117554   &5.33448e-14  \\
 13  &197.73176  &-1.4571297   &4.04592e-14  \\
 14  &197.74045  &-1.3452808   &2.70100e-14  \\
 15  &198.01117  &-1.2532794   &1.20219e-14  \\
 16  &198.00135  &-1.1785807   &3.28471e-14  \\
 17  &198.00404  &-1.4098294   &1.87843e-14  \\
 18  &197.92967  &-1.4700792   &3.34677e-14  \\
 19  &197.74718  &-1.4467673   &3.20633e-14  \\
 20  &197.68538  &-1.3966633   &2.40137e-14  \\
 21  &197.75135  &-1.3633463   &2.00829e-14  \\
 22  &198.01480  &-1.4067640   &2.44474e-14  \\
 23  &197.97641  &-1.3787678   &2.09356e-14  \\
 24  &197.77143  &-1.5433937   &2.80542e-14  \\
 25  &197.80647  &-1.5398204   &4.97257e-14  \\
 26  &197.89749  &-1.1571900   &2.98196e-14  \\
 27  &198.09163  &-1.3433972   &7.37440e-14  \\
 28  &198.00850  &-1.1652682   &5.30538e-14  \\
 29  &198.09833  &-1.3442155   &3.14757e-14  \\
 30  &197.82983  &-1.3427566   &1.38139e-14  \\
 31  &197.83316  &-1.3627567   &3.39411e-14  \\
 32  &197.83539  &-1.3827567   &1.35029e-14  \\
 33$^{*}$   &197.90652  &-1.3305344   &1.15971e-13  \\
 34$^{*}$   &198.11556  &-1.3692246   &1.31497e-14  \\
 35$^{*}$   &198.13638  &-1.2567236   &6.44595e-15  \\
 36$^{*}$   &198.13013  &-1.2358911   &7.99298e-15  \\
 37$^{*}$   &198.01554  &-1.5483976   &7.99298e-15  \\
 38$^{*}$   &197.74461  &-1.5254812   &4.89892e-15  \\
 39$^{*}$   &198.08012  &-1.2713117   &7.21947e-15  \\
 40  &197.99885  &-1.2879834   &5.73211e-15  \\
 41$^{*}$ &198.00302  &-1.3150664   &6.44595e-15  \\
 42  &197.62584  &-1.3150565   &1.38741e-13  \\
 43  &197.79672  &-1.2150684   &2.26699e-14  \\
 44  &197.81756  &-1.1859025   &2.09468e-14  \\
 45$^{*}$   &197.65711  &-1.2358936   &3.09405e-15  \\
 46$^{*}$   &197.67795  &-1.2150621   &2.06270e-15  \\
 47$^{*}$   &198.19683  &-1.3671319   &1.03135e-14  \\
 48  &198.05636  &-1.1827708   &2.38688e-14  \\
 49  &197.67990  &-1.4259208   &4.19095e-14  \\
\hline
 \end{tabular}
 \end{center}
 \tablenotetext{a}{Serial number for point sources. Sources with * on the number
 do not have counterparts in the 2XMMi catalog.}
 \tablenotetext{b}{Positions of point sources in the Equatorial J2000 coordinates (degrees).}
 \tablenotetext{c}{The 0.2--12.0 keV flux in units of erg s$^{-1}$ cm$^{-2}$.}
 \end{table}


\begin{thebibliography}{}
\bibitem[Abazajian et al.(2009)]{aba09}	
Abazajian, K. N.,  Adelman-McCarthy, J. K.,  Ag\"ueros, M. A.,  Allam, S. S.,  Allende P. C.,  An, D.,  Anderson, K. S. J.,  Anderson, S. F., et al. \ 2009,  ApJS, 182, 543
\bibitem[Afshordi et al.(2007)]{afshordi-2007} Afshordi, N., Lin, Y.-T.,
			       Nagai, D., \& Sanderson, A. J. R. 2007, MNRAS, 378, 293
\bibitem[Akahori \& Yoshikawa (2009)]{akahori-2009} Akahori, T. \& Yoshikawa, K. \ 2009, PASJ, submitted. Preprint: arXiv0909.5025
\bibitem[Anders \& Grevesse(1989)]{anders-1989}
 Anders, E., \& Grevesse, N. \ 1989, GeCoA, 53, 197
\bibitem[Atrio-Barandela et al.(2008)]{atrio-Barandela-2008}	
	Atrio-Barandela, F., Kashlinsky, A., Kocevski, D. \&  Ebeling, H. \ 
				      2008, ApJ, 675, L57
\bibitem[Bautz et al.(2009)]{bautz-2009}
Bautz, M. W., Miller, E. D., Sanders, J. S., Arnaud, K. A., Mushotzky, R. F., Porter, F. S., Hayashida, K., Henry, J. P., et al.  \ 2009, PASJ, 61, 1117
\bibitem[Bode et al.(2009)]{bode-2009} Bode, P., Ostriker, J. P. \& Vikhlinin, A. \  2009, ApJ, 700, 989
\bibitem[Borgani et al.(2005)]{borgani-2005}
Borgani, S., Finoguenov, A., Kay, S. T., Ponman, T. J., Springel, V., Tozzi, P., \& Voit, G. M. \ 2005, MNRAS, 361, 233
\bibitem[Broadhurst et al.(2005a)]{broadhurst-2005a} 
Broadhurst, T.,  Ben\'itez, N.,  Coe, D.,  Sharon, K.,  Zekser, K.,  White, R.,  Ford, H.,  Bouwens, R., et al. \ 2005,  ApJ, 621, 53
\bibitem[Broadhurst et al.(2005b)]{broadhurst-2005b}
Broadhurst, T., Takada, M., Umetsu, K., Kong, X., Arimoto, N., Chiba, M., \&  Futamase, T. \ 2005, ApJ, 619, L143
\bibitem[Broadhurst \& Barkana(2008)]{broadhurst-2008}
Broadhurst, T. \&  Barkana, R. \  2008, MNRAS, 390, 1647
\bibitem[Bullock et al.(2001)]{bullock-2001} 
Bullock, J. S., Kolatt, T. S., Sigad, Y., Somerville, R. S., Kravtsov, A. V., Klypin, A. A., Primack, J. R. \& Dekel, A. \ 2001, MNRAS, 321, 559
\bibitem[Cavaliere \& Fusco-Femiano (1976)]{cavaliere-1976} 
Cavaliere, A., \& Fusco-Femiano, R. \ 1976, A\&A, 49, 137
\bibitem[Cavaliere et al.(2009)]{cavaliere-2009}
Cavaliere, A., Lapi, A., \&  Fusco-Femiano, R. \  2009, ApJ, 698, 580 
\bibitem[Corless et al.(2009)]{corless-2009} 
Corless, V. L., King, L. J., \& Clowe, D. \  2009, MNRAS, 393, 1235
\bibitem[Croston et al.(2008)]{croston-2008}
Croston, J. H.,  Pratt, G. W.,  B\"ohringer, H.,  Arnaud, M.,  Pointecouteau, E.,  Ponman, T. J.,  Sanderson, A. J. R., Temple, R. F., et. al \ 2008, A\&A, 487, 431 
\bibitem[Dickey \& Lockman(1990)]{dickey-1990}
Dickey, J. M., \& Lockman, F. J. \ 1990, ARA\&A, 28, 215
\bibitem[Dolag et al.(2005)]{dolag-2005} 
Dolag, K.,  Vazza, F., Brunetti, G. \&  Tormen, G. \ 2005, MNRAS, 364, 753
\bibitem[Ettori \& Fabian(1998)]{ettori-1998} 
Ettori, S. \&  Fabian, A. C. \ 1998, MNRAS, 293, L33
\bibitem[Faltenbacher et al.(2007)]{faltenbacher-2007} 
Faltenbacher, A., Hoffman, Y., Gottl\"ober, S., \& Yepes, G. \  2007, MNRAS, 376, 1327
\bibitem[Fang et al.(2009)]{fang-2009} 
Fang, T., Humphrey, P., \& Buote, D. \ 2009, ApJ, 691, 1648
\bibitem[Fox \& Loeb(1997)]{fox-1997} 
Fox, D. C., \&  Loeb, A. \ 1997, 491, 459
\bibitem[Fujita et al.(2008)]{fujita-2008}
Fujita, Y.,  Tawa, N.,  Hayashida, K.,  Takizawa, M.,  Matsumoto, H.,  Okabe, N., \&  Reiprich, T. H. \ 2008, PASJ, 60, 343
\bibitem[Gavazzi(2005)]{gavazzi-2005} 
Gavazzi, R. \ 2005,A\&A, 443, 793
\bibitem[George et al.(2009)]{george-2009}
George, M. R., Fabian, A. C., Sanders, J. S., Young, A. J., \&  Russell, H. R. \ 2009, MNRAS, 395, 657
\bibitem[Hamana et al.(2003)]{hamana-2003} 
Hamana, T.,  Miyazaki, S.,  Shimasaku, K.,  Furusawa, H.,  Doi, M.,  Hamabe, M.,  Imi, K.,  Kimura, M., et al. \ 2003, ApJ, 597, 98
\bibitem[Hamana et al.(2009)]{hamana-2009} 
Hamana, T.,  Miyazaki, S.,  Kashikawa, N.,  Ellis, R. S.,  Massey, R. J.,  Refregier, A.,  Taylor, J.  E. \ 2009, PASJ, 61, 833
\bibitem[Henley \& Shelton(2008)]{henley-2008}
Henley, D. B., \& Shelton, R. L. \ 2008, ApJ, 676, 335
\bibitem[Ishisaki et al.(2007)]{ishisaki-2007}
Ishisaki, Y., Maeda, Y., Fujimoto, R., Ozaki, M., Ebisawa, K., Takahashi, T., Ueda, Y., Ogasaka, Y., et. al \ 2007, PASJ, 59, 113
\bibitem[Jeltema et al.(2008)]{jeltema-2008} 
Jeltema, T. E., Hallman, E. J., Burns, J. O., \& Motl, P. M. \ 2008, ApJ, 681, 167
\bibitem[King(1962)]{king-1962}
King, I. R. \ 1962, AJ, 67, 471
\bibitem[Komatsu \& Seljak(2001)]{komatsu-2001} 
Komatsu, E. \&  Seljak, U. \ 2001, MNRAS, 327, 1353
\bibitem[Komatsu et al.(2009)]{komatsu-2009} 
Komatsu, E., Dunkley, J., Nolta, M. R., Bennett, C. L., Gold, B., Hinshaw, G., Jarosik, N., Larson, D., et al. \ 2009, ApJS, 180, 330
\bibitem[Koyama et al.(2007)]{koyama-2007}
Koyama, K., Tsunemi, H.,; Dotani, T., Bautz, M. W., Hayashida, K., Tsuru, T. G.,  Matsumoto, H., Ogawara, Y., et al. \ 2007, PASJ, 59, 23
\bibitem[Kushino et al.(2002)]{kushino-2002}
Kushino, A., Ishisaki, Y., Morita, U., Yamasaki, N. Y., Ishida, M., Ohashi, T., \& Ueda, Y. \ 2002, PASJ, 54, 327
\bibitem[Lau, Kravtsov, \& Nagai (2009)]{lau-2009} 
Lau, E. T., Kravtsov, A. V., \& Nagai, D. \  2009, ApJ, 705, 1129
\bibitem[Lapi \& Cavaliere(2009)]{lapi-2009} 
Lapi, A. \&  Cavaliere, A. \  2009, ApJ, 692, 174
\bibitem[Lemze et al.(2009)]{lemze-2009} 
Lemze, D., Broadhurst, T., Rephaeli, Y., Barkana, R., \& Umetsu, K. \  2009, ApJ, 701, 1336
\bibitem[Limousin et al.(2007)]{limousin-2007} 
Limousin, M. \  2007, ApJ, 668, 643
\bibitem[Mitsuda et al.(2007)]{mitsuda-2007} 
Mitsuda, K.,  Bautz, M.,  Inoue, H.,  Kelley, R. L.,  Koyama, K.,  Kunieda, H.,  Makishima, K.  Ogawara, Y., et al. \ 2007, 59, S1
\bibitem[Miyazaki et al.(2002)]{miyazaki-2002} 
Miyazaki, S.,  Komiyama, Y.,  Sekiguchi, M.,  Okamura, S.,  Doi, M.,  Furusawa, H.,  Hamabe, M.,  Imi, K., et al. \ 2002, PASJ, 54, 833
\bibitem[Molnar et al.(2009)]{molnar-2009} 
Molnar, S. M., Hearn, N., Haiman, Z., Bryan, G., Evrard, A. E., \& Lake, G. \  2009, ApJ, 696, 1640
\bibitem[Molnar et al. (2010)]{molnar-2010} Molnar, S. et al. 2010, submitted to ApJL
\bibitem[Nagai \& Kravtsov(2003)]{nagai-2003} 
Nagai, D. \& Kravtsov, A. V. \ 2003, ApJ, 587, 514
\bibitem[Nagai, Vikhlinin, \& Kravtsov (2007)]{nagai-2007} 
Nagai, D., Vikhlinin, A., \& Kravtsov, A. V. \  2007, ApJ, 655, 98
\bibitem[Nakamura \& Suto(1997)]{nakamura-1997}  
Nakamura, T. T.,  \&  Suto, Y. \ 1997, Prog.Theor.Phys.,  97, 49
\bibitem[Nakazawa et al.(2009)]{nakazawa-2009}
Nakazawa, K., Sarazin, C. L., Kawaharada, M., Kitaguchi, T., Okuyama, S., Makishima, K., Kawano, N., Fukazawa, Y., et al.  \ 2009, PASJ, 61, 339
\bibitem[Navarro, Frenk, \& White(1996)]{navrro-1996}
Navarro, J. F.,  Frenk, C. S.,  \&  White, S. D. M. \ 1996, ApJ, 462, 563
\bibitem[Navarro, Frenk, \& White (1997)]{navrro-1997}
Navarro, J. F., Frenk, C. S., \& White, S. D. M. \ 1997, ApJ, 490, 493
\bibitem[Neto et al.(2007)]{neto-2007} 
Neto, A. F.,  Gao, L.,  Bett, P.,  Cole, S.,  Navarro, J. F.,  Frenk, C.  S.,  White, S. D. M.,  Springel, V., et al. \ 2007, MNRAS, 381, 1450
\bibitem[Oguri et al.(2005)]{oguri-2005} 
Oguri, M., Takada, M., Umetsu, K., \& Broadhurst, T. \ 2005, ApJ, 632, 841
\bibitem[Oguri et al.(2009)]{oguri-2009} 
Oguri, M., Hennawi, J. F., Gladders, M. D.,  Dahle, H., Natarajan, P., Dalal, N., Koester, B. P., Sharon, K. \& Bayliss, M. \ 2009, ApJ, 699, 1038
\bibitem[Okabe \& Umetsu(2008)]{okabe-2008} 
Okabe, N., \& Umetsu, K. \ 2008, PASJ, 60, 345 
\bibitem[Okabe et al.(2009)]{okabe-2009} 
Okabe, N., Takada, M., Umetsu, K., Futamase, T., \& Smith, G. P. \ 2009,  PASJ submitted. Preprint: arXiv:0903.1103 
\bibitem[Okabe et al.(2010)]{okabe-2010}
Okabe, N., Zhang, Y.,-Y., Finoguenov, A., Takada, M., Umetsu,  K., Smith,  G. P., \& Futamase, T. \ 2010,  ApJ submitted.
\bibitem[Ostriker et al.(2005)]{ostriker-2005}
Ostriker, J. P., Bode, P., \& Babul, A. \  2005, ApJ, 634, 964
\bibitem[Piffaretti et al.(2003)]{piffaretti-2003} 
Piffaretti, R., Jetzer, P., \& Schindler, S. \  2003, A\&A, 398, 41
\bibitem[Piffaretti  \& Valdarnini (2008)]{piffaretti-2008}  
Piffaretti, R., \& Valdarnini, R. \ 2008, A\&A, 491, 71
\bibitem[Peng et al.(2009)]{peng-2009}
Peng, E.-H., Andersson, K., Bautz, M. W., \&  Garmire, G. P. \ 2009, ApJ, 701, 1283
\bibitem[Ponman et al.(2003)]{ponman-2003}
Ponman, T. J., Sanderson, A. J. R., \&  Finoguenov, A. \ 2003, MNRAS, 343, 331
\bibitem[Pratt et al.(2006)]{pratt-2006}
Pratt, G. W., Arnaud, M., \&  Pointecouteau, E. \ 2006, A\&A, 446, 429
\bibitem[Pratt et al.(2007)]{pratt-2007}
Pratt, G. W., B\"ohringer, H., Croston, J. H., Arnaud, M., Borgani, S., Finoguenov, A., \& Temple, R. F. \ 2007, A\&A, 461, 71
\bibitem[Reiprich et al.(2009)]{reiprich-2009}
Reiprich, T. H., Hudson, D. S., Zhang, Y. -Y., Sato, K., Ishisaki, Y., Hoshino, A., Ohashi, T., Ota, N., et al. \ 2009, A\&A, 501, 899
\bibitem[Ryu et al.(2003)]{ryu-2003}	
Ryu, D., Kang, H., Hallman, E., \& Jones, T. W. \  2003, ApJ, 593, 599
\bibitem[Serlemitsos et al.(2007)]{serlemitsos-2007}    
Serlemitsos, P. J.,  Soong, Y., Chan. K., Okajima, T., Lehan, J. P., Maeda, Y., Itoh, K., Mori, H., et al. \ 2007, PASJ, 59, 9
\bibitem[Smith et al.(2001)]{smith-2001}    
Smith, R. K., Brickhouse, N. S., Liedahl, D. A., \& Raymond, J. C. \ 2001, ApJ, 556, L91
\bibitem[Snowden et al.(2008)]{snowden-2008}    
Snowden, S. L., Mushotzky, R. F.,  Kuntz, K. D., \&  Davis, D. S. \ 2008, A\&A, 478, 615
\bibitem[Spitzer(1962)]{spitzer-1962} 
Spitzer, L. Jr. \ 1962, Physics of Fully Ionized Gases (New York: Wiley)
\bibitem[Struble \& Rood(1999)]{struble-1999}
Struble, M. F., \& Rood, H. J. \ 1999, ApJS, 125, 35
\bibitem[Sunyaev \& Zeldovich(1972)]{sunyaev-1972} 
Sunyaev, R. A., \& Zeldovich, Y. B. \ 1972, Comments Astrophys. Space Phys., 4, 173
\bibitem[Takahashi et al.(2008)]{takahashi-2008} 
Takahashi, T.,  Kelley, R.,  Mitsuda, K.,  Kunieda, H.,  Petre, R.,  White, N.,  Dotani, T.,  Fujimoto, R., et al. \ 2008, Proc. SPIE, 7011, 70110O
\bibitem[Takizawa(1998)]{takizawa-1998} 
Takizawa, M. \  1998, ApJ, 509, 579
\bibitem[Takizawa (1999)]{takizawa-1999} 
Takizawa, M. \  1999, ApJ, 520, 514
\bibitem[Tawa et al.(2008)]{tawa-2008}
Tawa, N., Hayashida, K., Nagai, M., Nakamoto, H., Tsunemi, H., Yamaguchi, H., Ishisaki, Y., Miller, E. D., et al. \ 2008, PASJ, 60, 11
\bibitem[Taylor \& Navarro(2001)]{taylor-2001} 
Taylor, J. E., \& Navarro, J. F. \   2001, ApJ,  563, 483
\bibitem[Tozzi \& Norman(2001)]{tozzi-2001}
Tozzi P., \&  Norman, C. \ 2001, ApJ, 546, 63
\bibitem[Umetsu \& Broadhurst(2008)]{umetsu-2008}
Umetsu, K., \&  Broadhurst T. \ 2008, ApJ, 684, 177
\bibitem[Umetsu et al.(2009a)]{umetsu-2009a} 
Umetsu, K.,  Birkinshaw, M.,  Liu, G.,  Wu, J. P.,  Medezinski, E.,  Broadhurst, T.,  Lemze, D.,  Zitrin, A., et al. \ 2009, ApJ, 694, 1643
\bibitem[Umetsu et al.(2009b)]{umetsu-2009b} 
Umetsu, K., Medezinski, E., Broadhurst, T., Zitrin, A., Okabe, N., Hsieh, B. C., \& Molnar, S. M. \ 2009, ApJ submitted. Preprint:  arXiv0908.0069
\bibitem[Vazza et al.(2009)]{vazza-2009}
Vazza, F., Brunetti, G., Kritsuk, A., Wagner, R., Gheller, C., \& Norman, M. \ 2009, A\&A, 504, 33
\bibitem[Vikhlinin et al.(1999)]{vikhlinin-1999}
Vikhlinin, A., Forman, W., \& Jones, C. \ 1999, ApJ, 525, 47
\bibitem[Vikhlinin et al.(2006)]{vikhlinin-2006}
Vikhlinin, A., Kravtsov, A., Forman, W., Jones, C., Markevitch, M., Murray, S. S., \&  Van Speybroeck, L. \ 2006, ApJ, 640, 691
\bibitem[Vikhlinin et al.(2009)]{vikhlinin-2009}
Vikhlinin, A.,  Burenin, R. A.,  Ebeling, H.,  Forman, W. R.,  Hornstrup, A.,  Jones, C.,  Kravtsov, A. V.,  Murray, S. S., et al. \ 2009, ApJ, 692, 1033  
\bibitem[Voit et al.(2002)]{voit-2002}
Voit, G. M., Bryan, G. L., Balogh, M. L., \& Bower, R. G. \ 2002, ApJ, 601
\bibitem[Watson(2009)]{watson-2009}
Watson, M. G.,  Schr\"oder, A. C.,  Fyfe, D.,  Page, C. G., Lamer, G.,  Mateos, S.,  Pye, J.,  Sakano, M., et al. \ 2009, A\&A, 493, 339 
\bibitem[Zhang et al.(2006)]{zhang-2006} 
Zhang, Y.-Y., B\"ohringer, H., Finoguenov, A., Ikebe, Y., Matsushita, K., Schuecker, P., Guzzo, L., \& Collins, C. A. \ 2006, A\&A, 456, 55.
\bibitem[Zhang et al.(2007)]{zhang-2007} 
Zhang, Y.-Y., Finoguenov, A., B\"ohringer, H., Kneib, J.-P., Smith, G. P., Czoske, O. \& Soucail, G. \ 2007, A\&A, 467, 437
\bibitem[Zhang et al.(2008)]{zhang-2008} 
Zhang, Y.-Y., Finoguenov, A., B\"ohringer, H., Kneib, J.-P.,  Smith, G. P., Kneissl, R.,  Okabe, N., \& Dahle, H. \ 2008, A\&A, 482, 451
\bibitem[Zhang et al.(2010)]{zhang-2010} 
Zhang., Y-., Y., Okabe, N., Finoguenov, A., Graham, P. S., Piffaretti., R., Valdarnini, R., Babul, A., Evrard, A. E., et al. \ 2010 ApJ submitted. Preprint:  arXiv:1001.0780







\end{thebibliography}
\end{document}